%% file: MAGPI_h3h4Vasym_EnvQuench.tex
\documentclass[
  journal=pasa,
  manuscript=research-paper,
  year=2025,
]{cup-journal}

\usepackage{microtype,siunitx,booktabs,physics,multicol,slashbox,multirow,xcolor,orcidlink}

\usepackage{amssymb,xspace}
\newcommand{\kms}{\ensuremath{\mathrm{km\,s^{-1}}}\xspace}

\usepackage{subcaption}

\title{The MAGPI Survey: the subtle role of environment and not-so-subtle impact of generations of stars on galaxy dynamics}

\author{Caroline Foster\,\orcidlink{0000-0003-0247-1204}}
    \affiliation{School of Physics, University of New South Wales, Sydney, NSW 2052, Australia}
    \alsoaffiliation{ARC Centre of Excellence for All Sky Astrophysics in 3 Dimensions (ASTRO 3D)}
\author{Sabine Bellstedt\,\orcidlink{0000-0003-4169-9738}}
    \affiliation{International Centre for Radio Astronomy Research, The University of Western Australia, 35 Stirling Highway, Crawley WA 6009, Australia}
\author{Francesco D'Eugenio\,\orcidlink{0000-0003-2388-8172}}
    \affiliation{Kavli Institute for Cosmology, University of Cambridge, Madingley Road, Cambridge, CB3 0HA, United Kingdom}
    \alsoaffiliation{Cavendish Laboratory - Astrophysics Group, University of Cambridge, 19 JJ Thomson Avenue, Cambridge, CB3 0HE, United Kingdom}
    \alsoaffiliation{INAF -- Osservatorio Astronomico di Brera, via Brera 28, I-20121 Milano, Italy}
\author{Adriano Poci\,\orcidlink{0000-0002-5422-7441}}
    \affiliation{Sub-Department of Astrophysics, University of Oxford, Denys Wilkinson Building, Keble Road, Oxford OX1 3RH}
\author{Ryan Bagge\,\orcidlink{0009-0002-2753-3248}}
    \affiliation{School of Physics, University of New South Wales, Sydney, NSW 2052, Australia}
    \alsoaffiliation{ARC Centre of Excellence for All Sky Astrophysics in 3 Dimensions (ASTRO 3D)}
\author{Katherine Harborne\,\orcidlink{0000-0002-2043-7985}}
    \affiliation{International Centre for Radio Astronomy Research, The University of Western Australia, 35 Stirling Highway, Crawley WA 6009, Australia}
    \alsoaffiliation{ARC Centre of Excellence for All Sky Astrophysics in 3 Dimensions (ASTRO 3D)}
\author{Thomas Venville\, \orcidlink{0000-0003-0278-9933}}
    \affiliation{Research School of Astronomy and Astrophysics, Australian National University, Cotter Road, Weston Creek, ACT, 2611, Australia}
\author{J. Trevor Mendel\,\orcidlink{0000-0002-6327-9147}}
    \affiliation{Research School of Astronomy and Astrophysics, Australian National University, Cotter Road, Weston Creek, ACT, 2611, Australia}
    \alsoaffiliation{ARC Centre of Excellence for All Sky Astrophysics in 3 Dimensions (ASTRO 3D)}
\author{Claudia Del P. Lagos\,\orcidlink{0000-0003-3021-8564}}
    \affiliation{International Centre for Radio Astronomy Research, The University of Western Australia, 35 Stirling Highway, Crawley WA 6009, Australia}
    \alsoaffiliation{ARC Centre of Excellence for All Sky Astrophysics in 3 Dimensions (ASTRO 3D)}
\author{Emily Wisnioski\,\orcidlink{0000-0003-1657-7878}}
    \affiliation{Research School of Astronomy and Astrophysics, Australian National University, Cotter Road, Weston Creek, ACT, 2611, Australia}
    \alsoaffiliation{ARC Centre of Excellence for All Sky Astrophysics in 3 Dimensions (ASTRO 3D)}
\author{Tania M. Barone\,\orcidlink{0000-0002-2784-564}}
    \affiliation{Centre for Astrophysics and Supercomputing, Swinburne University of Technology, Hawthorn, VIC 3122}
    \alsoaffiliation{School of Physics, University of New South Wales, Sydney, NSW 2052, Australia}
    \alsoaffiliation{ARC Centre of Excellence for All Sky Astrophysics in 3 Dimensions (ASTRO 3D)}
\author{Andrew J. Battisti\orcidlink{0000-0003-4569-2285}}
    \affiliation{International Centre for Radio Astronomy Research, The University of Western Australia, 35 Stirling Highway, Crawley WA 6009, Australia}
    \alsoaffiliation{Research School of Astronomy and Astrophysics, Australian National University, Cotter Road, Weston Creek, ACT, 2611, Australia}
    \alsoaffiliation{ARC Centre of Excellence for All Sky Astrophysics in 3 Dimensions (ASTRO 3D)}
\author{Stefania Barsanti\,\orcidlink{0000-0002-9332-5386}}
    \affiliation{Research School of Astronomy and Astrophysics, Australian National University, Cotter Road, Weston Creek, ACT, 2611, Australia}
    \alsoaffiliation{ARC Centre of Excellence for All Sky Astrophysics in 3 Dimensions (ASTRO 3D)}
    \alsoaffiliation{Sydney Institute for Astronomy, School of Physics, A28, The University of Sydney, NSW 2006, Australia}
\author{Sarah Brough\,\orcidlink{0000-0002-9796-1363}}
    \affiliation{School of Physics, University of New South Wales, Sydney, NSW 2052, Australia}
    \alsoaffiliation{ARC Centre of Excellence for All Sky Astrophysics in 3 Dimensions (ASTRO 3D)}
\author{Scott M. Croom\,\orcidlink{0000-0003-2880-9197}}
    \affiliation{Sydney Institute for Astronomy, School of Physics, A28, The University of Sydney, NSW 2006, Australia}
    \alsoaffiliation{ARC Centre of Excellence for All Sky Astrophysics in 3 Dimensions (ASTRO 3D)}
\author{Caro Derkenne\,         \orcidlink{0000-0003-3474-3542}}
    \affiliation{School of Mathematical and Physical Sciences, Macquarie University, NSW 2109, Australia}
    \alsoaffiliation{ARC Centre of Excellence for All Sky Astrophysics in 3 Dimensions (ASTRO 3D)}
\author{Lucas C. Kimmig\,\orcidlink{0009-0006-8337-8712}}
    \affiliation{Universitäts-Sternwarte München, Fakultät für Physik, Ludwig-Maximilians Universität, Scheinerstr. 1, D-81679 München, Germany}
\author{Anilkumar Mailvaganam}
    \affiliation{School of Mathematical and Physical Sciences, Macquarie University, NSW 2109, Australia}
    \alsoaffiliation{Macquarie University Astrophysics and Space Technologies Research Centre, Sydney, NSW 2109, Australia}
    \alsoaffiliation{ARC Centre of Excellence for All Sky Astrophysics in 3 Dimensions (ASTRO 3D)}
\author{Rhea-Silvia Remus\,\orcidlink{0009-0008-9260-7278}}
    \affiliation{Universitäts-Sternwarte München, Fakultät für Physik, Ludwig-Maximilians Universität, Scheinerstr. 1, D-81679 München, Germany}
\author{Gauri Sharma\,\orcidlink{ 0000-0002-6070-2851}}
    \affiliation{Observatoire Astronomique de Strasbourg, Université de Strasbourg, CNRS UMR 7550, F-67000 Strasbourg, France}
\author{Sarah M. Sweet\,\orcidlink{0000-0002-1576-2505}}
    \affiliation{School of Mathematics and Physics, University of Queensland, Brisbane, QLD 4072, Australia}
    \alsoaffiliation{ARC Centre of Excellence for All Sky Astrophysics in 3 Dimensions (ASTRO 3D)}
\author{Sabine Thater,\orcidlink{0000-0003-1820-2041}}
    \affiliation{Department of Astrophysics, University of Vienna, Türkenschanzstraße 17, 1180 Vienna, Austria}
\author{Lucas M. Valenzuela\,\orcidlink{0000-0002-7972-9675}}
    \affiliation{Universitäts-Sternwarte München, Fakultät für Physik, Ludwig-Maximilians Universität, Scheinerstr. 1, D-81679 München, Germany}
\author{Jesse van de Sande\,\orcidlink{0000-0003-2552-0021}}
    \affiliation{School of Physics, University of New South Wales, Sydney, NSW 2052, Australia}
    \alsoaffiliation{ARC Centre of Excellence for All Sky Astrophysics in 3 Dimensions (ASTRO 3D)}
\author{Sam P. Vaughan\,\orcidlink{0000-0003-2265-7727}}
    \affiliation{School of Mathematical and Physical Sciences, Macquarie University, NSW 2109, Australia}
    \alsoaffiliation{ARC Centre of Excellence for All Sky Astrophysics in 3 Dimensions (ASTRO 3D)}
\author{Bodo Ziegler\,\orcidlink{0000-0003-2856-1080}}
    \affiliation{Department of Astrophysics, University of Vienna, Türkenschanzstraße 17, 1180 Vienna, Austria}


\usepackage{hyperref} 
\hypersetup{colorlinks,citecolor=blue,linkcolor=blue,urlcolor=blue}

\begin{document}

\begin{abstract}
The stellar age and mass of galaxies have been suggested as the primary determinants for the dynamical state of galaxies, with environment seemingly playing no or only a very minor role.
We use a sample of 77 galaxies at intermediate redshift ($z\sim0.3$) in the Middle-Ages Galaxies Properties with Integral field spectroscopy (MAGPI) Survey to study the subtle impact of environment on galaxy dynamics. We use a combination of statistical techniques (simple and partial correlations and principal component analysis) to isolate the contribution of environment on galaxy dynamics, while explicitly accounting for known factors such as stellar age, star formation histories and stellar masses. We consider these dynamical parameters: high-order kinematics of the line-of-sight velocity distribution (parametrised by the Gauss-Hermite coefficients $h_3$ and $h_4$), kinematic asymmetries $V_{\rm asym}$ derived using {\sc kinemetry} and the observational spin parameter proxy $\lambda_{R_e}$. Of these, the mean $h_4$ is the only parameter found to have a significant correlation with environment as parametrised by group dynamical mass. This correlation exists even after accounting for age and stellar mass trends.
We also find that satellite and central galaxies exhibit distinct dynamical behaviours, suggesting they are dynamically distinct classes.
Finally, we confirm that variations in the spin parameter $\lambda_{R_e}$ are most strongly (anti-)correlated with age as seen in local studies, and show that this dependence is well-established by $z\sim0.3$.
\end{abstract}

\section{Introduction}\label{sec:introduction}

The morphological and dynamical mix of galaxies has been observed and theorised to evolve across cosmic time \citep[e.g.][]{Abraham01,Gnedin03,Bezanson18,Lagos22,Cavanagh23}. In particular, distributions of the observational proxy for the spin parameter as defined in \citet{Emsellem07} measured using mock observations of various hydro-dynamic simulations confirm the expectation that stars in galaxies spin down on average as a population with cosmic time \citep[e.g.][]{Lagos18b,Schulze18,Foster21}.

Observations that older stellar populations are on hotter orbits than younger stars has been observed in the Milky Way and other galaxies \citep{Quirk19,Poci19,Shetty20,Poci21,Foster23}, suggesting either that stars were born dynamically hotter in the past or that older stellar populations have had more time to dynamically heat through interactions than recently formed stars. This suggests that stellar dynamics either reflect the dynamical conditions at the time of formation, the cumulative effect of dynamical heating post-formation or a combination of both \citep{Leaman17}.

Integral field spectroscopy (IFS) studies of samples of galaxies in the local Universe ($z\lesssim0.1$) have revealed that low spin or non-rotating galaxies are preferentially found in denser environments: the so-called kinematic morphology-density relation \citep[e.g.][]{Cappellari11b,DEugenio13,Houghton13,Fogarty14,Scott14,Cappellari16}. This trend with environmental density was initially interpreted as a sign of nurture. In other words, the environment was thought to be responsible for the spin down of galaxies by providing more opportunities for merging and dynamical heating.

However, recent studies have questioned this line of reasoning, suggesting that environment only plays a secondary role in setting the spin of galaxies \citep[e.g.][]{Brough17,Greene17,Veale17,Wang20,Rutherford21,vandeSande21}. Instead, galaxy spin has been shown to anti-correlate primarily with age, rather than mass or environment \citep{Croom24}. \citet{Croom24} do however find a weak residual correlation with environment when considering only the most massive galaxies. Similarly, \citet{Vaughan24} found that environment and local density are not major contributors to a galaxy's rotational state once accounting for mass, size, star formation rate and apparent ellipticity. 
\citet{MunozLopez24} recently found no evidence for environmental impact on spin in a sample of intermediate redshift galaxies in the COSMOS fields. Even the spin alignment of galaxies with large scale filament structures seems to be primarily driven by bulge mass (\citealt{Barsanti22}; Barsanti et al. in prep).

The diminishing role of environment as a key driver for the dynamical state of galaxies seems at odds with accepted wisdom around hierarchical merging, which suggests environment and the incidence of mergers play a crucial role in galaxy evolution. A possibly helpful way to consider these results is that environment plays a secondary role through its impact on the hierarchical growth of mass and star formation quenching. Indeed, how can we reconcile recent studies suggesting a lack of statistical dependence of the proportion of merging galaxies on environmental density at $z\sim0$ \citep{vanDokkum99,alonso12,Sureshkumar24} with theoretical predictions that most group and cluster centrals exhibit tidal features \citep{Khalid24} and the observed higher merging fractions in higher galaxy over-densities at high redshift \citep{Shibuya24}?

The above results suggest that careful selection of the observational properties under scrutiny and to be contrasted with environment plays a crucial role and additional tracers to the traditional observational spin proxy parameter ($\lambda_{R_e}$) are needed to detect the more subtle environmental impact on dynamics. Environment metrics themselves also vary broadly, with some probing immediate environment, local density, global environment, or the broader large scale structure in which galaxies are embedded \citep[e.g.][]{Cooper05,Colberg08,Muldrew12}. The choice of environment metric, shape of the probed volume and depth of the survey used may also impact observational studies and the detectability of environment trends.

Using Schwarzschild orbit-superposition modelling with triaxial potentials for the Sydney-Australian-Astronomical-Observatory Multi-object Integral-Field Spectrograph (SAMI) kinematic maps, \citet{Santucci23} have shown that both environment and stellar mass impact the fraction of stars on dynamically hot vs. warm orbits in low mass galaxies. This suggests that more detailed dynamical measurements may hold the key to teasing out the role of environment in altering the dynamics of galaxies.

Other recent studies suggest that environment does play a role in setting other properties of galaxies, sometimes even when accounting for other factors explicitly. For example, environmental density affects the quenching of galaxies, which impacts the measured age and star formation history (SFH), with lower mass galaxies being more susceptible to environmental quenching \citep[e.g.][but see \citealt{Darvish16}]{Schaefer17,Tan22,Wang22,Oxland24,RomeroGomez24}. 

Most statistically large studies of stellar kinematics using IFS have been limited to $z<0.1$. Beyond this redshift, statistical studies were restricted to single-slit \citep[LEGA-C][$z\sim1$]{vanderWel16}, and IFS observations were limited to single objects \citep[e.g.][]{PerezGonzalez24}, clusters \citep[e.g.][]{Mahler18} or well-known fields \citep[e.g. HUDF, GOODS-S and COSMOS,][]{Guerou17,MunozLopez24}.
The Middle-Ages Galaxies Properties with Integral field spectroscopy (MAGPI\footnote{Based on observations obtained at the Very Large Telescope (VLT) of the European Southern Observatory (ESO), Paranal, Chile (ESO program ID 1104.B-0536).}) Survey aims to address this by pushing the redshift limit for statistical studies of stellar kinematics out to $z\sim0.3$ using a representative sample of $>150$ spatially resolved galaxies (including a mix of isolated, centrals and satellites) covering a range of morphologies (see also \citealt{Foster21}; Foster et al. in prep). For our purposes, isolated galaxies are defined as those galaxies that are not members of an identified group or cluster.

In this work, we leverage the MAGPI sample and turn to a range of dynamical parameters (in addition to $\lambda_{R_e}$) in an attempt to detect the possibly subtle impact of environment on the stellar dynamics of galaxies at intermediate redshift ($z\sim0.3$), while carefully controlling for previously identified confounding factors such as stellar mass, age and SFH.

This paper is structured as follows: the data and sample selection are presented in \S \ref{sec:data}. Our analysis and results can be found in \S \ref{sec:analysis}. A discussion and our conclusions are presented in \S \ref{sec:discussion} and \S \ref{sec:conclusions}, respectively.

We assume a $\Lambda$CDM universe with $H_0$ = 70 km s$^{-1}$ Mpc$^{-1}$, $\Omega_{\rm M}$ = 0.3 and $\Omega_\Lambda$ = 0.7. We assume a \citep{Chabrier03} initial mass function (IMF) throughout. All magnitudes are in the AB magnitude system \citep{Oke1983}.

\section{Data}\label{sec:data}


\begin{figure*}
\includegraphics[width=18cm]{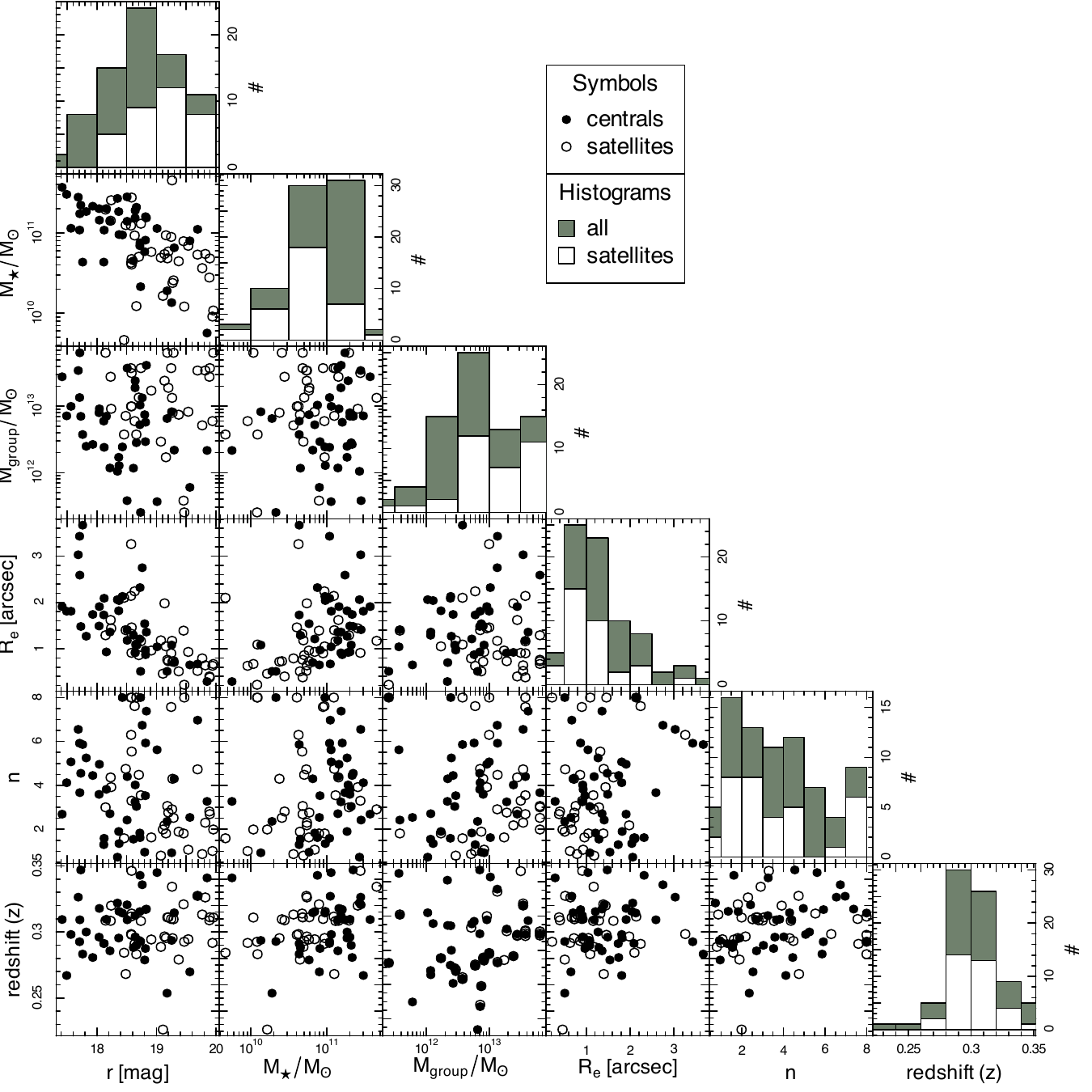}
\caption{Corner plot comparing a selection of basic properties for our selected galaxies and showing their distributions (histogram in right most panel of each row). Centrals and satellite galaxies are shown as filled and hollow symbols, respectively. Green and hollow histograms at the end of each row represent all and satellites, respectively. Shown properties are (left to right and top to bottom): magnitude ($r$), stellar mass ($M_{\star}$), group mass ($M_{\rm group}$), effective radius ($R_e$), S\'ersic index (n) and redshift (z).} \label{fig:selection_hist}
\end{figure*}

\subsection{The MAGPI Survey}\label{sec:MAGPI}

The MAGPI Survey\footnote{\url{http://magpisurvey.org/}} is a Very Large Telescope (VLT) Multi-Unit Spectroscopic Explorer (MUSE) large programme targeting 60 massive ($M_\star > 7 \times 10^{10}M_\odot$) central galaxies at intermediate redshift ($0.25 < z < 0.35$, primaries) and their immediate environment. The MAGPI sample selection was based on the Galaxy and Mass Assembly (GAMA) survey \citep{Driver11}. Importantly, the sample was designed to span a broad range of halo masses (i.e. $11.35 \le \log_{10} ( M_{\rm halo}) \le 15.35$) and galaxy colours to ensure representation of all galaxy types.
We refer to \citet{Foster21} and Mendel et al. (in preparation) for details of the MAGPI survey strategy, sample selection and science goals, along with a description of the data reduction and quality assessment. We provide a brief summary of the data reduction steps for completeness. 

The raw MUSE data cubes are reduced using the {\sc pymusepipe}\footnote{\url{https://github.com/emsellem/pymusepipe}} interface for the ESO MUSE pipeline \citep{Weilbacher12,Weilbacher20}. The pipeline is used to perform the standard bias and overscan subtraction, flat-fielding, wavelength calibration, telluric correction and cube reconstruction steps. The Zurich Atmosphere Purge \citep[ZAP,][]{Soto16} package is used to improve background sky subtraction.

Since the MUSE cubes represent the deepest images available and to ensure uniformity in detecting targets, galaxies and other objects are detected directly on the white light MUSE image using the {\sc ProFound r} package \citep{Robotham18}. Segmentation maps are created within {\sc ProFound} to define the edges of every detected source.

The segmentation maps are used to cut the main MUSE data cube into ``minicubes'' for every detected object. The software package {\sc mpdaf}\footnote{\url{https://github.com/musevlt/mpdaf}} is used to produce minicubes and synthetic Sloan Digital Sky Survey (SDSS) filter images in $g$, $r$ and $i$.

Basic structural parameters (e.g. effective radius $R_e$, photometric position angle $PA_{\rm phot}$, Sersi\'c indices $n$, etc) in all three synthetic bands $g$, $r$ and $i$ are obtained using both {\sc ProFound} and {\sc GalFit} \citep{Peng02,Peng10}. {\sc ProFound} also provides magnitudes in $g$, $r$ and $i$.

For each identified object, a redshift is measured using MARZ \citep{Hinton16}, including both an initial automated estimate and subsequent visual inspection. We use a modified template set provided by M. Fossati \footnote{\url{https://matteofox.github.io/Marz/}}, which includes additional higher resolution and high-redshift templates that are well matched to the variety of objects detected in the MAGPI data.

\subsection{Parameter derivation}\label{sec:paramderive}

\subsubsection{Stellar masses and star formation histories}

\begin{figure}
\includegraphics[width=8.5cm]{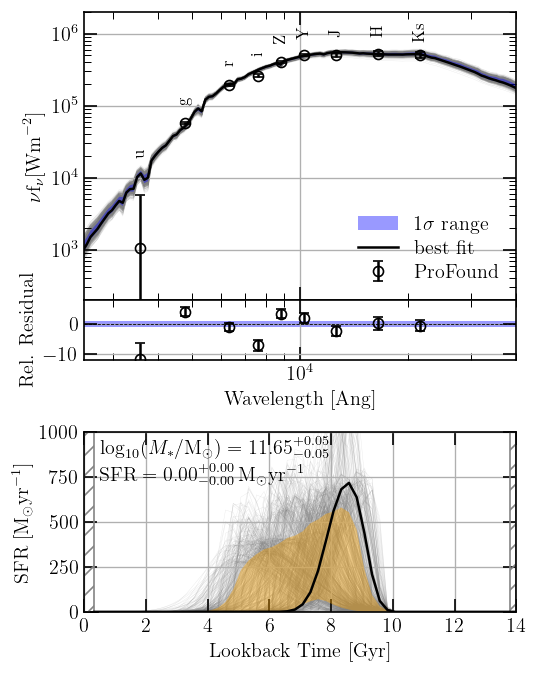}
\caption{Illustration of the {\sc ProSpect} output for MAGPI2307228105. 
Top: {\sc ProSpect} spectral energy distribution fit to the observed broadband magnitudes (best fit shown in black, thin posterior distribution shown in grey, and the 1$\sigma$ range shown in blue), with the residual fit shown below.
Bottom: Corresponding star formation history. Grey lines show the thinned Monte Carlo Markov Chain posterior distribution, black line shows the posterior mode, with the orange shaded region showing the 1$\sigma$ range. The galaxy has $M_\star/M\odot=10^{11.65}$. }\label{Fig:prospect}
\end{figure}

Following the methodology developed for the GAMA survey \citep{Bellstedt20b, Driver22}, the {\sc ProSpect} spectral energy distribution fitting code \citep{Robotham20} is used to derive stellar masses and SFHs based on broad-band photometry in 9 bands ($u$-$K_s$). Forced photometry based on the segmentation maps is derived using images from the Kilo-Degree Survey \citep[KiDS,][]{deJong17} and VISTA Kilo-degree Infrared Galaxy \citep[VIKING,][]{Edge13} that are pixel-matched to the MAGPI minicubes. We model the star formation history assuming a skewed normal distribution truncated in the early universe to a null star formation rate. We assume a linearly evolving metallicity with time, along with a \citet{Chabrier03} initial mass function (IMF), \citet{Bruzual03} simple stellar population spectra, stellar and nebular attenuation law by \citet{Charlot00} and dust re-emission library of far-infrared templates presented by \citet{Dale14}. 
\textsc{ProSpect} is fitted to each galaxy using MCMC with 10,000 steps. Global parameters such as the stellar mass, star formation rate, and age are computed for each step in the chain, allowing for the extraction of a median value and a 1$\sigma$ range to capture the overall value and uncertainty. These median values are used for all galaxy properties throughout this work (rather than the values as derived by the single best-fitting step from the MCMC chain). 
More detail on the methodology can be found in \citet{Bellstedt20b}. Figure \ref{Fig:prospect} shows an example fit.

\subsubsection{Stellar kinematics}

\begin{figure*}
\includegraphics[width=18cm]{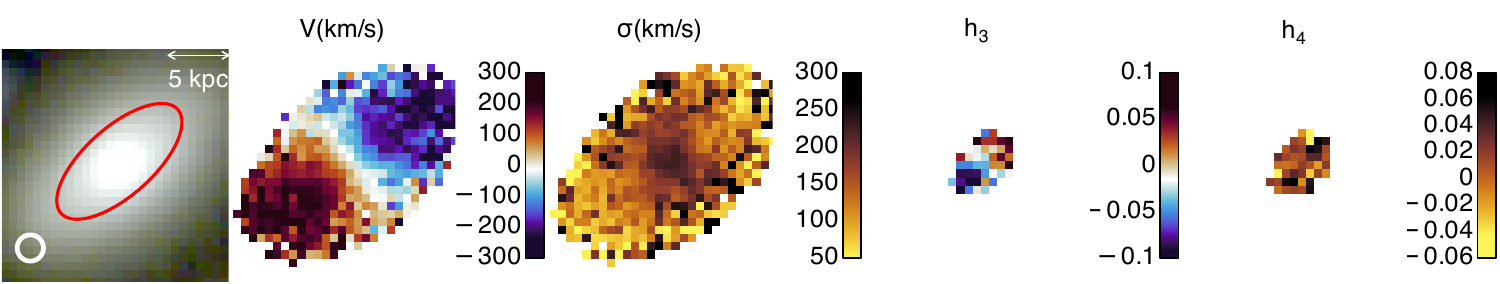}
\caption{Synthetic $g$, $r$, $i$ MUSE image for MAGPI2307228105 (left) with PSF (FWHM) illustrated as a white circle in the lower corner and physical scale provided on the top right. The effective radius is shown with a red ellipse. To the right of the image and on the same scale, the four measured higher-order kinematic moments maps (from left to right): velocity ($V$), velocity dispersion ($\sigma$), $h_3$ and $h_4$, as labelled. The comparatively more stringent selection criterion for higher order kinematics lead to less spatially extended $h_3$ and $h_4$ maps than those of $V$ and $\sigma$. This galaxy has dynamical parameters $\lambda_{R_e}=0.68$, $\rho_{V-h_3}=-0.79$, $p_{V-h_3}<0.001$, $\mu_{h_4}=0.015$.}\label{Fig:kinmaps}
\end{figure*}

Stellar kinematic velocity ($V$) and velocity dispersion ($\sigma$) maps are obtained through spectral fitting using the Galaxy Integral field unit Spectroscopy Tool \footnote{\url{https://pypi.org/project/gistPipeline/}} \citep[{\sc gist},][]{Bittner19}, which is a wrapper for the penalised pixel cross-correlation fitting {\sc python} package ({\sc pPXF}, \citealt{Cappellari04,Cappellari17}). Our method is similar to that used in the SAMI Galaxy Survey with 4 moments as described in \citet{vandeSande17a,Croom21}. First, {\sc pPXF} is fit to concentric elliptical annular bins with photometric position angle and axis ratio to determine an optimal subset of templates. The optimal sets of templates for the respective bin and immediately adjacent bins are then combined and used to fit individual spaxels within the bin (see \citealt{Foster21} and \citealt{DEugenio23a} for further detail). We set the keyword \texttt{bias} to a function of the empirical signal-to-noise ratio ($S/N$, see~\ref{sec:ppxfbias}). Only individual spaxels with a continuum signal-to-noise ratio above 3 per pixel are fit after masking spectral regions of possible nebular emission and strong skylines using a series of stellar templates from the IndoUS stellar template library \citep{Valdes04}. For the velocity and velocity dispersion maps, we select spaxels with a velocity dispersion uncertainty $\sigma_{\rm err}<25 + 0.1\sigma$ km s$^{-1}$ following \citet{vandeSande17b} and only keep galaxies with an 85 percent minimum fill fraction within $1R_e$ for $V$ and $\sigma$. We select a threshold of $S/N>15$ per pixel for the $h_3$ and $h_4$ maps. A more stringent threshold leads to a smaller dataset of higher fidelity data points. We note that our results are qualitatively robust against a range of $10<S/N<25$ thresholds for $h_3$ and $h_4$. Example kinematic maps are shown in Figure \ref{Fig:kinmaps}.

Visual classifications of the stellar kinematics into galaxies that show obvious rotation (OR) and those that do not (NOR) are derived based on 11 independent classifications. We use the mode of the Bayesian posterior for high confidence (probability of classification being correct $P>0.98$) classifications from Foster et al. (in preparation).

\subsubsection{Spin parameter proxy}

We use the spin parameter proxy values produced by \citet{Derkenne24}. Briefly, the stellar kinematic maps are used to compute the observational spin parameter proxy $\lambda_{r}$ as defined by \citet{Emsellem07}:
\begin{equation}
\lambda_r=\frac{\sum^{N}_{i=1} F_i R_i\abs{v_i}}{\sum^{N}_{i=1}F_i R_i \sqrt{v_i^2+\sigma_i^2}},
\end{equation}
where $F_i$, $R_i$, $v_i$ and $\sigma_i$ are the flux, galactocentric radius, recession velocity corrected for the systemic velocity and the velocity dispersion measured in the $i^{\rm th}$ spaxel within an aperture of size $r$, respectively. Here, $R_i$ represents the length of the semi-major axis of the ellipse if the i$^{\rm th}$ spaxel rather than a projected circular apertures. We use $r=R_e$ (i.e. one effective radius) elliptical apertures, closely following the methodology of \citet{FraserMcKelvie22} as described in \citet{Derkenne24}. Finally, a seeing and aperture correction is applied using the code of \citet{Harborne20}.

\subsubsection{Kinematic asymmetries}

We use {\sc kinemetry} \citep{Krajnovic06} to decompose the stellar velocity maps of our galaxies using a Fourier Series along concentric ellipses. For each ellipse with position angle PA and axis ratio $q=b/a$, where $a$ and $b$ are the semi-major and -minor axes, respectively, the velocity at a given azimuth angle with respect to the semi-major axis $\theta$ can be approximated using:
\begin{equation}
K(a,\theta)=A_0+\sum^{m=N}_{m=1} A_m \sin(m\theta) + B_m \cos(m\theta),
\end{equation}
where $A_0$ is the zeroth harmonic term and $A_m$ and $B_m$ are the $m$th harmonic terms. The kinemetric fits are described in detail in \citet{Bagge23}, to which we refer the reader for more detail. Based on the fits, $k_m$ parameters and $V_{\rm asym}$ are computed as follows:
\begin{equation}
k_m=\sqrt{A^2_m+B^2_m};
V_{\rm asym}=\frac{k_2+k_3+k_4+k_5}{4S_{0.5}},
\end{equation}
where $S_{0.5}=\sqrt{0.5V_{\rm rot}^2+\sigma^2}$, is a proxy for dynamical mass in units of km s$^{-1}$ that is robust across galaxy morphological types \citep[see][for a detailed justification and calculations]{Bagge24}.

\subsubsection{Stellar ages}
\DeclareSIUnit[per-mode=reciprocal-positive-first]{\year}{yr}
We use mass-weighted stellar population ages on integrated spectra (i.e. co-adding all spaxels within the relevant segment). Stellar population parameters are derived using the full spectral fitting code {\sc pPXF} to fit simple stellar population models from the E-MILES library of \citet{Vazdekis16} with a Chabrier IMF \citep{Chabrier03} with the BaSTI isochrones \citep{Hidalgo18}. The fit automatically masks emission lines from ionised gas and other spurious spectral pixels such as under-subtracted sky emission lines. We include only the templates within the safe ranges \citep{Vazdekis16} of age \((t)\) and total metallicity \((\textrm{[M/H]})\), which is approximately between \(-2.0\) to \(+0.4\) dex for metallicity, and a youngest age of \(\SI{0.1}{\giga\year}\). We also impose a maximum age set to the age of the Universe at the redshift of each galaxy. These models are derived using an \([\alpha/\mathrm{Fe}]\) abundance of the Solar neighbourhood. As we are only interested in the relative average ages within our sample (observed at roughly a single epoch), we opt for the single-\([\alpha/\mathrm{Fe}]\) E-MILES models over MILES, in favour of the extended wavelength coverage. More information on the stellar populations for MAGPI will be provided in Poci et al. (in preparation).

\subsubsection{Group masses}

Following the methodology of \citet{Knobel09} and \citet{Robotham11}, environmental metrics for MAGPI are calculated using {\sc parliament}\footnote{A commonly used collective noun for a group of magpies is a ``parliament''.} (Harborne et al. in preparation; Bravo et al. in preparation), a friend-of-friend grouping algorithm run on available redshifts within the MAGPI field-of-view. In this work, we will make use of the group mass proxy (assuming a multiplicative factor of $A=10$, see \citealt{Robotham11}), which is a dynamical mass derived based on the group velocity dispersion and Virial Theorem arguments. These group masses are improved from those presented in \citet{Foster21} thanks to the inclusion of MAGPI redshifts. While the methodology has been tested on the GAMA survey, we note that the MAGPI sample is not complete and thus group masses may be underestimated should some members of the group lie outside the probed field-of-view.

Group members are then classified as ``centrals'' if they dominate their group in the $i$-band magnitude (i.e. they are the brightest member). Other group members are deemed ``satellites''.

We also explored potential trends with the distance to the nearest neighbour within the group and number of group members, but did not find a significant correlation with those environmental metrics.

\subsection{Parameter selection}\label{sec:paramselect}

In order to contrast our results with similar published studies, we include the spin parameter $\lambda_{R_e}$ as a measure of overall rotational vs. pressure support. We include other dynamical properties such as the stellar kinematic asymmetry as measured from kinemetry, thought to be an indicator of recent interactions. A further 2 ``subtle'' parameters are derived based on the higher-order Gauss-Hermite polynomial third $h_3$ and fourth $h_4$ moment of the line-of-sight velocity distribution (LOSVD). The first is the Spearman rank correlation coefficient of the velocity vs. $h_3$ maps ($\rho_{v-h_3}$) as an indication of the prominence of disc-like orbits in the LOSVD \citep[e.g.][]{Naab14,vandeSande17a}. The second is simply the inverse-error-weighted mean $h_4$ for valid spaxels within $1R_e$ ($\mu_{h_4}$) as resulting from the presence of hot orbits in the wings of the LOSVD.

We note that $\mu_{h_4}$ contrasts with that presented in \citet[][whose parameter we will henceforth refer to as $H_4$]{DEugenio23a} in that here $h_4$ is measured on individual spaxels before averaging instead of directly fitting the aperture spectrum (also see \ref{sec:appendix1} for a direct comparison). \citet{DEugenio23a} also used a toy model to show (their figure 2) that $H_4$ (integrated) strongly correlates with $\mu_{h_4}$ (local). \ref{sec:appendix1} shows a similar trend is present in our MAGPI data, albeit with significant observational scatter. We choose the average local $\mu_{h_4}$ instead of the integrated $H_4$ in order to minimise the possible impact of artificial line broadening by rotation due to beam smearing the LOSVD, although we note that our conclusions are unchanged for $H_4$ (see \ref{sec:appendix1}). A positive $h_4$ is associated with broader wings and a more central peak than a standard Gaussian distribution \citep[e.g.][]{Bender94}. 

A brief discussion on the potential impact of beam smearing on $\rho_{V-h_3}$ and $\mu_{h_4}$ is included in \S\ref{sec:pcorrs}, where we mention testing our results while explicitly accounting for seeing (FWHM) as a confounding parameter.

To parametrise the star formation histories in our galaxies, we select 3 parameters. The first is the mass-weighted stellar population age as measured using {\sc pPXF}. We also consider the lookback time of the peak ($\mu_{\rm SFH}$) and the difference in lookback time that brackets the formation of 10-90 percent of the stars ($\delta_{\rm SFH}$) of the parametrised SFHs derived with {\sc ProSpect}. For galaxies where the SFH has yet to peak, $\mu_{\rm SFH}$ values are set to 0.
Together, these parameters are used to quantify the peak and extent of the SFHs of our targets. 
We exclude $\mu_{\rm SFH}$, $\delta_{\rm SFH}$ and stellar mass values inferred for MAGPI2307197200 because for this galaxy alone the lack of far-infrared data during the SED fitting resulted in a large dust content being fitted, thereby overestimating the attenuation and pushing the fit to a very high mass for the respective $r$-band. This overestimate in the dust was identified through a comparison of the measured far-infrared photometry for this galaxy from the GAMA survey. This was the only such outlier identified. Generally the stellar masses inferred for MAGPI galaxies are aligned with those of the GAMA survey where available\footnote{We note that MAGPI masses tend to scatter lower, as the higher resolution of MAGPI means that some GAMA galaxies are sometimes resolved into multiple galaxies.} \citep{Bellstedt20b, Driver22}. 

We use group mass ($M_{\rm group}$) to parametrise the environment of our targets.
We note that the group masses are qualitatively in line with those of the GAMA Survey, using the same method, but not to be compared directly quantitatively due to probing different volumes and depths. Finally, we separately consider central (i.e. brightest group member in $i$-band) and satellite galaxies.

The possible impact of stellar mass ($M_\star$) on dynamics is also considered.

\begin{figure}
\includegraphics[width=8.5cm]{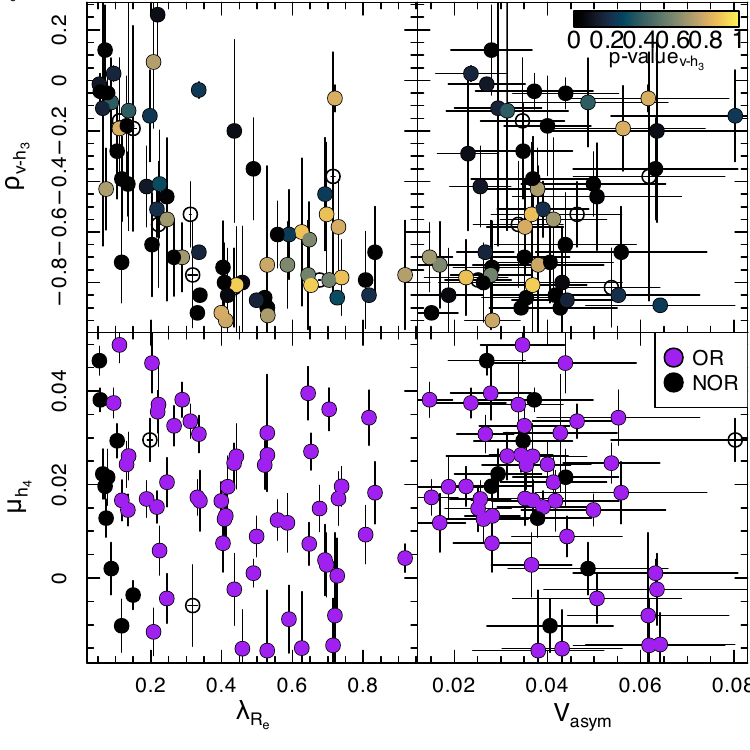}
\caption{Comparison of higher-order kinematic moment parameters used in this work against the spin parameter $\lambda_{R_e}$ and stellar kinematic asymmetries measured with kinemetry $V_{\rm asym, stars}$. When available, data are colour-coded according to the p-value of the $v-h_3$ in the top row, or whether the galaxies were visually identified with obvious rotation (OR, purple) or without obvious rotation (NOR, black) in the bottom row according to visual classifications as per Foster et al. (in prep.). Lighter symbols in the top row indicate galaxies for which the $V-h_3$ anti-correlation is of lower statistical significance.}\label{fig:dyn_tracers}
\end{figure}


\subsection{Sample selection}\label{sec:sampleselection}

\begin{figure}
\includegraphics[width=8.5cm]{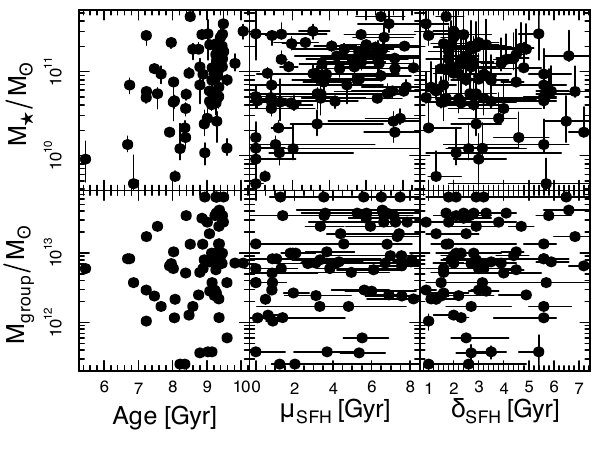}
\caption{Comparison of star formation history proxies (mass-weighted age, SFH peak, SFH duration $\delta_{\rm SFH}$) with stellar mass ($M_\star$) and group mass ($M_{\rm group}$) confirming known trends are present in our data. Uncertainties are shown whenever available.}\label{fig:ageSFH_env}
\end{figure}

To minimise the impact of possible confounding factors and high measurement uncertainties, we select a high-quality sample of MAGPI galaxies as follows. First, we apply an $r$-band magnitude cut of $r<20$ mag and limit our study to those galaxies near or at the MAGPI nominal redshift by selecting galaxies at $0.2 < z < 0.4$. To ensure reliable kinematic proxies (i.e. $V_{\rm asym}$, $\lambda_{R_e}$, $\rho_{v-h_3}$, $\mu_{h_4}$) we select galaxies where the stellar kinematics covering fraction is $>0.85$ within $1R_e$ and only include spin values within the physical range $0<\lambda_{R_e}<1$ after seeing correction.
Figure \ref{fig:selection_hist} shows the magnitude, stellar mass, group mass, size, S\'ersic index and redshift distributions of our selected sample. This final curated sample contains 77 galaxies. This final sample of 77 galaxies includes 41 centrals, 34 satellites and 2 isolated galaxies (MAGPI1527067139 and MAGPI2306197198).

Figure \ref{fig:dyn_tracers} illustrates how the dynamical parameters derived based on higher-order kinematic moments (i.e. $\rho_{v-h_3}$, $\mu_{h_4}$) compare with $\lambda_{R_e}$ and $V_{\rm asym}$. As expected, there is a clear anti-correlation between $\lambda_{R_e}$ and $\rho_{v-h_3}$, suggesting that high spin galaxies exhibit more disc-like orbits than lower spin galaxies. We do not find an obvious correlation between $V_{\rm asym}$ and $\rho_{v-h_3}$ or $\lambda_{R_e}$ and $\mu_{h_4}$. Only a small fraction of galaxies exhibit slightly negative mean $h_4$. 

Figure \ref{fig:ageSFH_env} shows the underlying correlations between SFH proxies, stellar mass and group mass present in our data. As expected higher mass galaxies are typically older with SFH peaks further in the past than their lower mass counterparts \citep[e.g.][]{Gallazzi05,Deng15}. Older galaxies also tend to be found preferentially in richer environments (i.e. higher mass groups) and vice versa as already seen in other studies \citep[e.g.][]{Wolf07,Tiwari20}.

\section{Analysis and results}\label{sec:analysis}

We employ a range of statistical techniques to analyse what factors contribute to the dynamics of galaxies. We present our results via each of these techniques. The first is a simple correlation (\S \ref{sec:simplecorr}), we next attempt to reduce the parameter space using a principal component analysis (\S \ref{sec:PCA}) and then follow up remaining parameters using partial correlations (\S \ref{sec:pcorrs}).

\subsection{Spearman correlation}\label{sec:simplecorr}

\begin{table*}
\caption{Compiled Spearman rank correlation coefficients $\rho_{\rm Spearman}$ and respective p-values for each pairs of considered parameters. The number of galaxies where both considered parameters are available (i.e. complete cases, $N_{\rm CC}$) for each test is given. Significant correlations (i.e. p-value $< 0.02$) are highlighted in bold. Considered dynamical parameters are the strength of the anti-correlation between $V$ and $h_3$ ($\rho_{V-h_3}$), mean $h_4$ ($\mu_{h_4}$), stellar kinematic asymmetry ($V_{\rm asym}$) and the spin parameter proxy ($\lambda_{R_e}$) compared with intrinsic properties: stellar mass ($M_{\star}$, column 1), environment as parameterised through group mass ($M_{\rm group}$, column 2), mass weighted stellar age (Age, column 3), the lookback time of the peak of the star formation history ($\mu_{\rm SFH}$, column 4); and the duration of the SFH ($\delta_{\rm SFH}$).}
\input{table_spearman}
\label{table:corrs}
\end{table*}

\begin{figure*}
\includegraphics[width=18cm]{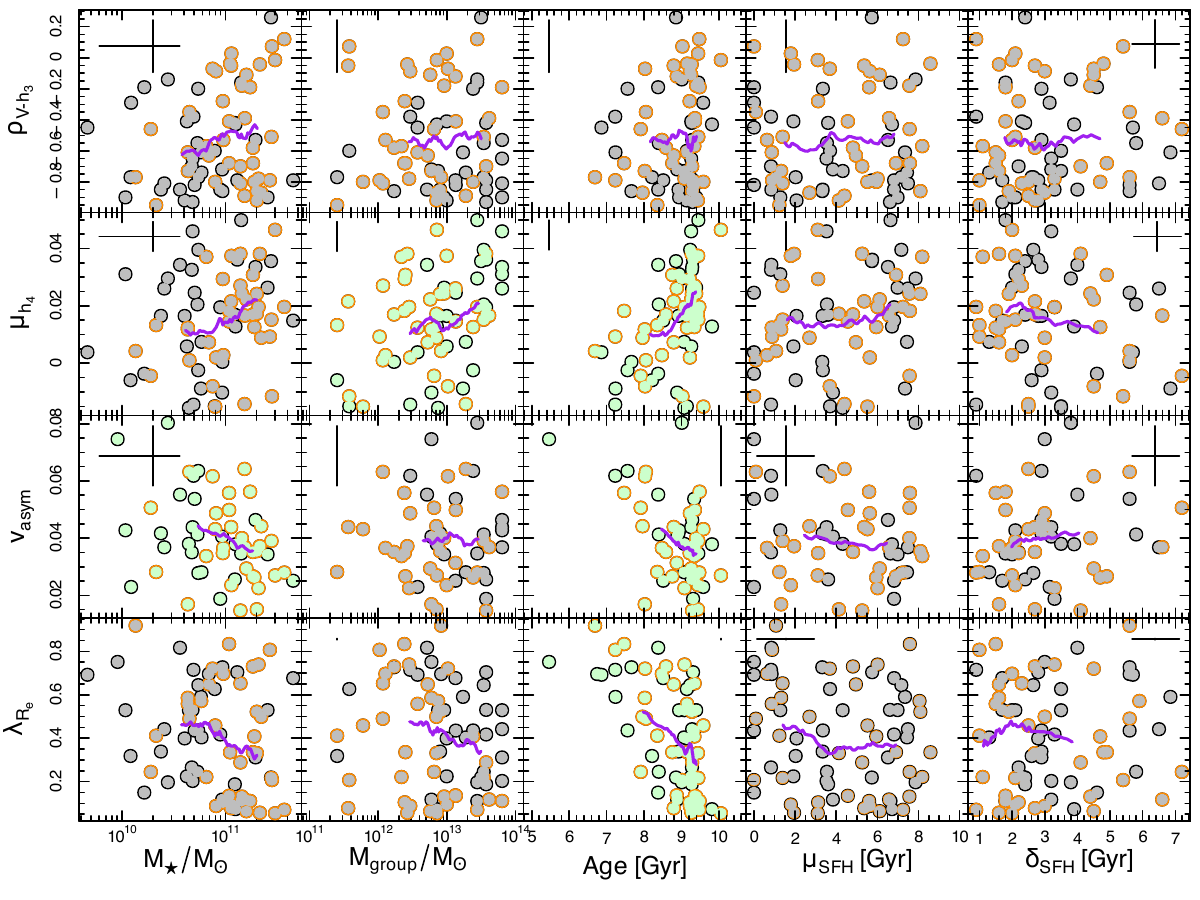}
\caption{Identifying correlations between dynamical parameters ($\rho_{V-h_3}$, $\mu_{h_4}$, $V_{\rm asym,stars}$ and $\lambda_{R_e}$) and stellar mass ($M_\star$), group mass ($M_{\rm group}$), mass-weighted stellar age (Age), lookback time of the SFH peak and the 10-90 percent SFH ($\delta_{\rm SFH}$). Grey and green symbols are used when data are correlated at the $<98$ (no significant correlation detected) and $\ge98$ (i.e. significant correlation) percent confidence, respectively. Centrals are shown as symbols with orange outlines and satellites with black outlines. Results of the Spearman rank correlation analysis are given in Table \ref{table:corrs}. Median uncertainties are shown in each panel whenever available. Rolling means of bin size 30 are shown in purple to guide the eye.}\label{fig:corrs}
\end{figure*}

Firstly, we want to determine which of our dynamical parameters correlate with other considered parameters. In what follows, we will refer to parameters as simply being correlated whether they are positively or negatively (i.e. anti) correlated. We perform a simple Spearman rank correlation test on all relevant pairs of variables (see Figure \ref{fig:corrs} and Table \ref{table:corrs}).
The Spearman rank method first ranks the data in each parameter and compares the rankings rather than the original values. As such, this methodology may be used to detect the presence of non-linear correlations so long as they are monotonic in nature.

As we are looking for the potentially subtle residual role of environment on galaxy dynamics, we do also consider weak, but significant correlations. In what follows, we consider a correlation statistically significant if the $p$-value is below a threshold of $p\le0.02$. In other words, we are willing to wrongly identify a correlation at most 2 percent of the time (i.e. 98 percent confidence). We choose this threshold over a more stringent one in order to be inclusive of potentially relevant parameters to be considered in subsequent analysis, while also allowing for the exclusion of less relevant parameters.
We note that even a weak (or subtle) correlation ($\abs{\rho}\lesssim0.5$) may be statistically significant and that a strong correlation ($\abs{\rho}\sim1$) may not be statistically significant, though the latter is unlikely for sufficiently large samples. 

With these criteria, we examine Fig. \ref{fig:corrs} and Table \ref{table:corrs}, and find a number of weak but statistically significant correlations ($M_{\rm group}$ and Age vs. $\mu_{h_4}$; $M_\star$ and Age vs. $V_{\rm asym}$; and Age vs. $\lambda_{R_e}$) as well as weak marginally significant correlations (i.e. $p$-value$<0.05$, $M_\star$ and $\mu_{\rm SFH}$ vs. $\mu_{h_4}$, and $M_\star$ vs. $\lambda_{R_e}$).

Importantly, there is no statistically significant or marginal correlation between $\rho_{V-h_3}$ or $\delta_{\rm SFH}$ and any of the other considered parameters. The lack of correlation with the latter may be due to the large uncertainties on this parameter, which reflect the inherent difficulties with inferring star formation histories. This suggests we may be able to simplify our analysis by reducing the number of parameters considered by removing $\rho_{V-h_3}$ or $\delta_{\rm SFH}$. We explore this further in \S \ref{sec:PCA}.

Apart from the parameters involving $M_{\rm group}$ and $M_\star$, other parameter pairs do not show a visible difference in the distribution of the central or satellite galaxies.

\subsection{Principal component analysis}\label{sec:PCA}

\begin{figure*}
\includegraphics[width=8cm]{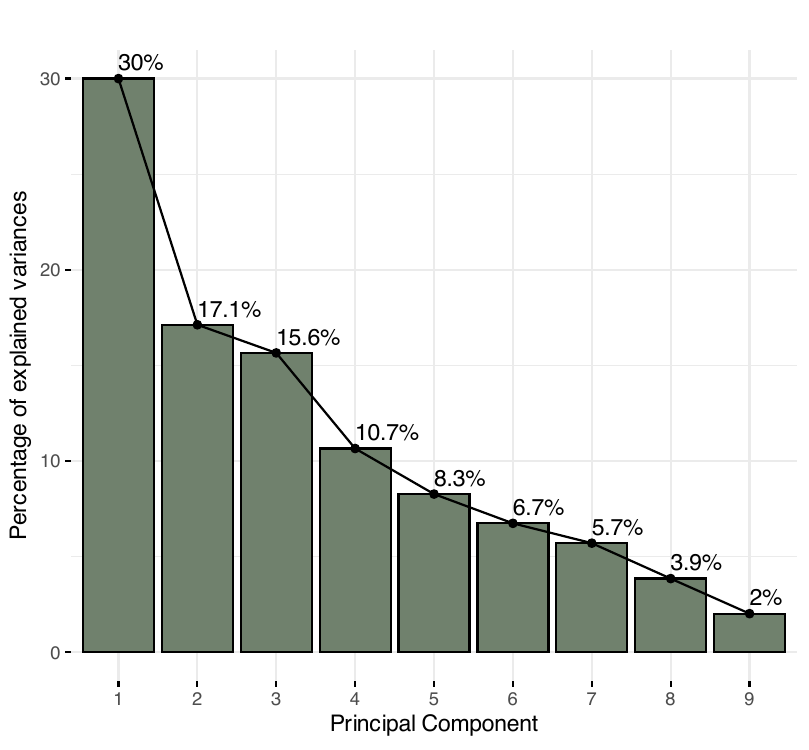}
\includegraphics[width=9cm]{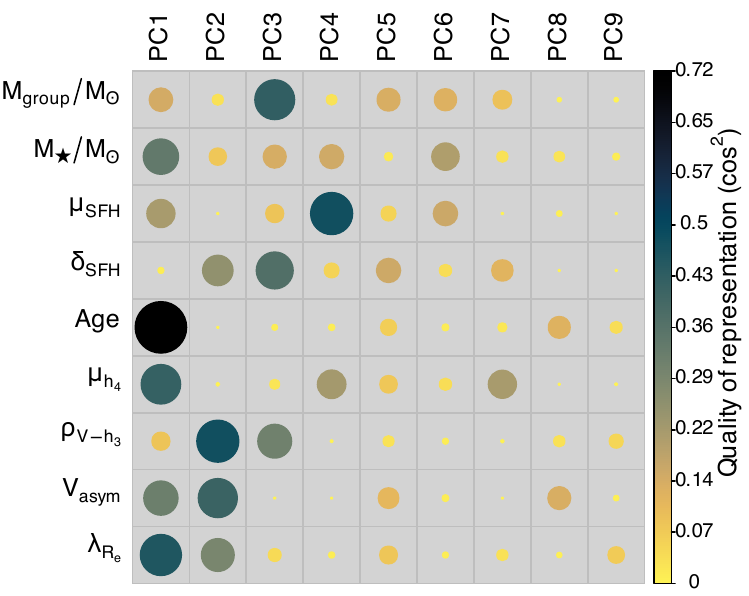}
\caption{Overview of the principal component analysis results. The first 5 components explain $>80$ percent of the variance in the data (left). The quality of representation of the data (cos$^2$) for each parameter and principal component is illustrated with circles with colour and size both representing cos$^2$ (right). Most of the variance in the data (PC1, 30 percent) is dominated by age, with $M_\star$, $\mu_{h_4}$ and $\lambda_{R_e}$ also showing significant quality of representation. A total of 50 galaxies where all parameters are available (i.e. complete cases) are included in this analysis. Similarly, PC2 (representing 17.1 percent of the variance) suggests $\rho_{V-h_3}$, $V_{\rm asym}$ and $\lambda_{R_e}$ are co-variant. The $\delta_{\rm SFH}$ parameter dominates has its highest quality of representation in PC3 (15.6 percent of the variance) along with $M_{\rm group}$ and $\rho_{V-h_3}$, but little co-variance with other dynamical parameters. $\mu_{\rm SFH}$ dominates PC4.}\label{fig:pca}
\end{figure*}

We use principal component analysis \citep[see][for a review]{Jollife16} to study the directions in our chosen nine-dimensional parameter space along which most of the variance is seen in our dataset. The first principal component captures the direction of the largest variance, with subsequent components being orthogonal to all others and explaining decreasing fractions of the variance in the sample. A short principal component explains very little of the variance and thus some of the co-variant parameters may be rejected to help reduce the dimensionality of the problem. We are also interested in finding and discarding parameters that do not correlate with other considered parameters in a statistically meaningful way. We note that a parameter with a large amount of noise due to large measurement uncertainties may show up as a contributor to an early principal component. Such a noisy parameter will generally not be correlated with other parameters.

In practice, we use the R package {\sc pracma}\footnote{\url{https://cran.r-project.org/package=pracma}} to perform a principal component analysis on the 51 galaxies for which all fitted parameters are available. The left panel of Figure \ref{fig:pca} shows the percentage of the variance explained by each of the principal components. The first 6 components (out of the 9 fitted) are sufficient for explaining $\sim90$ percent of the variance in the data. The quality of representation for each parameter is commonly parametrised by the sum of the square values of the cosine (cos$^2$) of the ``angle from the right triangle made with the origin, the observation, and its projection'' \citep{Abdi10} on a principal component. In other words, the sum of the distances to individual measurements along the principal component. Components with large associated cos$^2$ values contribute a commensurately large portion to the variance along that component. Individual parameters with the largest cos$^2$ values within a component are best represented by that component. For each parameter, cos$^2$ is shown on the right panel of Figure \ref{fig:pca}, which visually illustrates which parameters tend to vary together in the dataset.

We examine the data using PCA and find that the first five components account for over 80 percent of the variance in the data, with the first four exhibiting the highest quality of representation. The first principal component (PC1, explaining 30 percent of the variance) is dominated by stellar age, with $M_\star$, $\mu_{h_4}$, $V_{\rm asym}$ and $\lambda_{R_e}$ showing the highest qualities of representation. Indeed, stellar age is co-variant with most other parameters studied (except $\delta_{\rm SFH}$) and those with the highest quality of representation are thus relevant parameters to control for. Since there is no significant correlation of $\delta_{\rm SFH}$ with any of the dynamical parameters under study (Figure \ref{fig:corrs}), we infer that the duration of the SFH of galaxies is either too uncertain/noisy or a redundant parameter in our analysis. We thus choose to exclude this parameter from subsequent analysis, though we will return to it in \S\ref{sec:discussion}.

PC2 (17.1 percent of the variance) suggests that the dynamical parameters $\rho_{V-h_3}$, $V_{\rm asym}$ and $\lambda_{R_e}$ are co-variant. Given this and the fact that $\rho_{V-h_3}$ did not show significant correlation with any of the other parameters in Figure \ref{fig:corrs}, we also exclude this parameter from further analysis as a redundant parameter (i.e. its variance is already captured by other considered parameters).

\subsection{Spearman partial correlation}\label{sec:pcorrs}

\begin{figure}
\includegraphics[width=8.5cm]{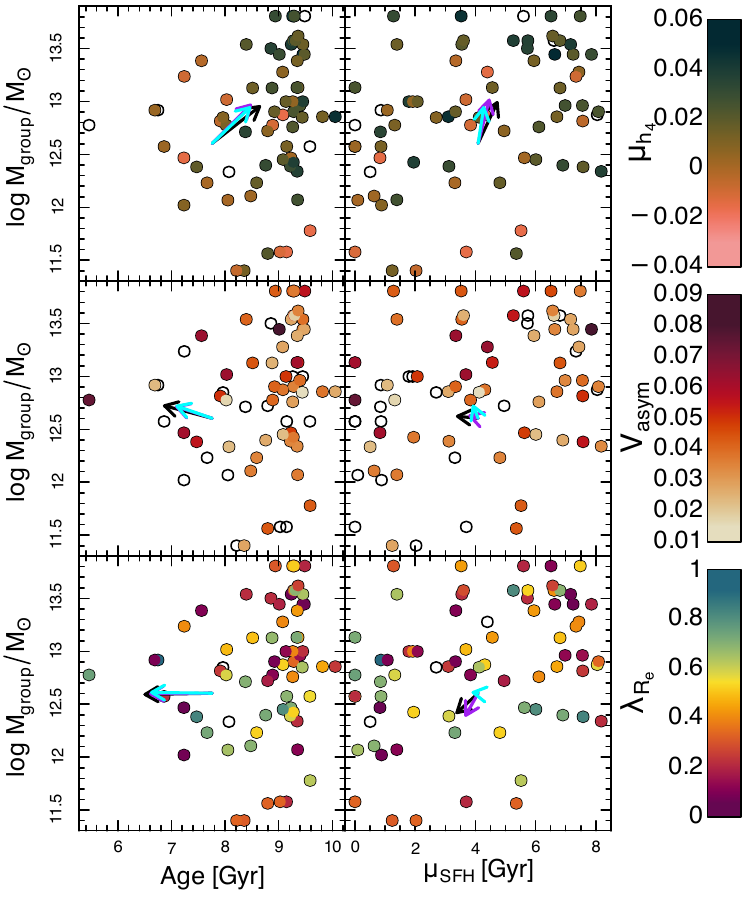}
\caption{Partial correlation analysis for $z_i=\mu_{h_4}$ (top), $z_i=V_{\rm asym}$ (middle) and $z_i=\lambda_{R_e}$ (bottom) as a function of group mass ($y=M_{\rm group}$) and stellar age ($x={\rm Age}$, left) or $x=\mu_{\rm SFH}$ (right). Hollow symbols are used for missing values. Black arrows show the direction and strength of the partial correlation for the parameters on the respective axes (i.e. $x={\rm Age},y=M_{\rm group},\textbf{Z}=\{z_i\}$). Purple arrow shows the partial correlation while simultaneously accounting for the plotted variables and stellar mass (i.e. $x={\rm Age},y=M_{\rm group},\textbf{Z}=\{z_i,M_\star\}$). The cyan arrows show the partial correlations while accounting for stellar mass, age, group mass and $\mu_{\rm SFH}$ simultaneously (i.e. $x={\rm Age},y=M_{\rm group},\textbf{Z}=\{z_i,\mu_{\rm SFH},M_\star\}$). Partial correlation coefficients and p-values are recorded in Table \ref{table:pcorrs}. While some of the variance in the $M_{\rm group}$ vs. age or $\mu_{\rm SFH}$ plot is accounted for by other variables, there remains a significant correlation with for $z_i=\mu_{h_4}$. This indicates that  $M_{\rm group}$, age and $\mu_{\rm SFH}$ all individually contribute to the variance in $\mu_{h_4}$.}
\label{fig:allpcorrs}
\end{figure}



\begin{table*}
\caption{Compiled Spearman rank partial correlation coefficients $\rho$ and respective significance ($p$-values) for groups of considered parameters in this work. The number of galaxies where all considered parameters are available (i.e. complete cases, $N_{\rm CC}$) for each test is given (column 6). Significant correlations (i.e. $p\le 0.02$) are highlighted in bold. Considered dynamical parameters are the mean $h_4$ ($\mu_{h_4}$), stellar kinematic asymmetry ($V_{\rm asym}$) and spin parameter proxy ($\lambda_{R_e}$). Thus, $x = {\rm Age}$ or $\mu_{\rm SFH}$, $y=M_{\rm group}$, and $\textbf{Z}$ is a subset of $\{z_i, M_\star, \mu_{\rm SFH}\}$ for dynamical parameters $z_i\in\{\mu_{h_4},V_{\rm asym}, \lambda_{R_e}\}$ (as labelled in column 1) in Eq. \ref{eq:pcorr}. These dynamical parameters are compared with intrinsic properties: group mass ($M_{\rm group}$, column 2), mass weighted stellar age (Age, column 3), the lookback time of the peak of the SFH ($\mu_{\rm SFH}$, column 4); and stellar mass ($M_\star$, column 5). Each row corresponds to a separate test with excluded parameters marked with dashes ``-''.}
\input{table_pcorrs}
\label{table:pcorrs}
\end{table*}

The use of partial correlation analysis \citep[e.g.][]{Kendall42,Lawrance76} has gained in popularity for this type of astronomy studies in recent years (e.g. \citealt{Baker22,Baker23,Koller24,Croom24,Bluck24}). Partial correlation analysis allows one to measure the correlation coefficient while accounting for other variables. 

In practice, we employ the Spearman rank method as implemented in the R package {\sc ppcor}\footnote{\url{https://cran.r-project.org/package=ppcor}} \citep{kim15} to perform partial correlation analyses and tease out the impact of environment on our considered dynamical parameters, while explicitly accounting for known important factors (namely stellar mass, age and/or SFH parameters).

Mathematically, the Spearman rank partial correlation coefficient for variable $z_i$ while accounting for variables $x$, $y$, and $\textbf{Z} \setminus\{z_i\}$ (where $\textbf{Z}$ represents a vector of multiple variables, i.e. $\textbf{Z}=\{z_1,z_2,...\}$) can be written as follows for any $z_i \in \textbf{Z}$: 
\begin{equation}\label{eq:pcorr}
\rho_{xy | \textbf{Z}} = \frac{\rho_{xy | \textbf{Z} \setminus \{z_i\}} - \rho_{xz_i |\textbf{Z} \setminus\{z_i\}}\rho_{z_iy \textbf{Z} \setminus\{z_i\}}}{\sqrt{1-\rho_{xz_i|\textbf{Z}\setminus\{z_i\}}}\sqrt{1-\rho_{z_iY|\textbf{Z}\setminus\{z_i\}}}},
\end{equation}
where $x = $ Age or $\mu_{\rm SFH}$, $y=M_{\rm group}$ and $\textbf{Z}$ is a subset of $\{z_i, M_\star, {\rm Age}, \mu_{\rm SFH}\}$ for $z_i\in\{\mu_{h_4},V_{\rm asym}, \lambda_{R_e}\}$. The statistical notation `$\setminus$' in Equation \ref{eq:pcorr} signifies exclusion of the nominated parameter(s) from the set.

We note that for the Spearman rank method, an assumption that trends are monotonic is made. As such, the Spearman rank method makes fewer assumptions about the form of the correlation than Pearson (which assumes a linear trend), and is therefore more generally applicable when the correlation is not known a priori. However, we note that should a trend not be monotonic, this may not be captured adequately with this methodology. Also, thanks to the ranking process, results are immune from our choice of logging the axis and our choice of unit.

We now look at the results of the partial correlation analysis. Figure \ref{fig:allpcorrs} illustrates how the remaining dynamical parameters (i.e., $\mu_{h_4}$, $V_{\rm asym}$ and $\lambda_{R_e}$) vary as a function of $M_{\rm group}$ and stellar age or $\mu_{\rm SFH}$. Missing values are shown as open symbols and are not used in the relevant analyses. Results from the partial correlation analysis are illustrated with coloured arrows, where the length of the arrow in each direction is proportional to the correlation coefficient for that variable once accounting for the others. Partial correlation coefficients and respective $p$-values are listed in Table \ref{table:pcorrs}.

\begin{figure}
\includegraphics[width=8.5cm]{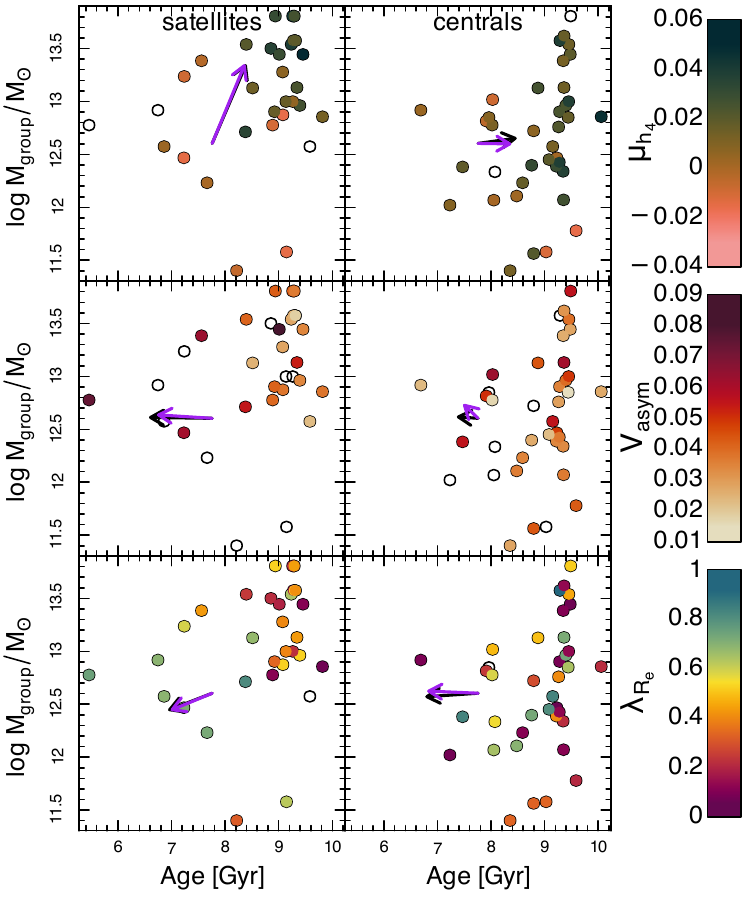}
\caption{Same as left panels of Figure \ref{fig:allpcorrs}, but for satellites (left) and centrals (right) separately. The only significant partial correlation with group mass is that of $\mu_{h_4}$ for satellites (top left panel, refer to relevant $p$-values quoted in Table \ref{table:pcorrs}).}
\label{fig:satcen}
\end{figure}

Environment (i.e. $M_{\rm group}$) significantly correlates with $\mu_{h_4}$, even when accounting for stellar mass and age ($\rho=0.30$, $p=0.016$) in the sample. The dependence on group mass is not statistically significant when the analysis is performed on the centrals only. This implies that it is mainly the dynamics of satellite galaxies that are driving the trends with environment as parametrised by $M_{\rm group}$ in our sample. To illustrate the different behaviour between the satellites and centrals, Figure \ref{fig:satcen} shows the two samples separately. We note that the number statistics are quite low for these sub-samples, especially when controlling for multiple parameters. Despite this, values tabulated in Table \ref{table:pcorrs} show there an even more significant partial correlation of $\mu_{h_4}$ with group mass when considering satellites only.

Importantly, the impact of environment in the whole sample is only ever significant for the $\mu_{h_4}$ parameter (Figure \ref{fig:allpcorrs}), where stellar age also shows significant partial correlations whenever considered for the whole sample.
$V_{\rm asym}$ is mainly impacted by stellar age (and perhaps marginally stellar mass, $p$-value$=0.021$ when not simultaneously accounting for stellar age), with no significant partial correlations for $M_{\rm group}$ or $\mu_{\rm SFH}$.

Figure \ref{fig:allpcorrs} and Table \ref{table:pcorrs} show that age most strongly correlates with $\lambda_{R_e}$, even after accounting for all other considered variables, with no statistically significant partial correlations with other considered parameters (see Table \ref{table:pcorrs}). When considering centrals and satellites separately, the partial correlation with age is only significant for the centrals sub-sample.

In each row of Figure \ref{fig:allpcorrs}, the cyan arrows show the partial correlation after accounting for all other relevant parameters. Trends with $\mu_{h_4}$, $V_{\rm asym}$ and $\lambda_{R_e}$ for centrals often diverge (in direction and strength) from those with satellites (refer to Fig. \ref{fig:satcen} and Table \ref{table:pcorrs}). Indeed, when separating satellites and centrals, the only significant partial correlation with $\mu_{h_4}$ is $M_{\rm group}$. In contrast, when the same separation is made for $\lambda_{R_e}$, the only significant partial correlation is that with stellar age in centrals only. We note however the challenges in detecting correlations in a relatively small sample, especially after subdividing into satellite and central sub-samples and accounting for multiple parameters as done here.

The impact of beam smearing on the higher order kinematic moments is not straightforward to infer. In order to quantify the potential impact of beam smearing on our $\mu_{h_4}$ results, we have repeated the analysis with the FWHM in $r$-band as one of the parameters within $\textbf{Z}$. There were no statistically significant partial correlations with FWHM for any of the considered parameters. We consider that beam smearing effects do not account for the trends seen with $\mu_{h_4}$.

\section{Discussion}\label{sec:discussion}

In this work, we consider the complex interplay between stellar dynamical parameters ($\rho_{V-h_3}$, $\mu_{h_4}$, $V_{\rm asym}$ and $\lambda_{R_e}$), stellar mass ($M_\star$), environment as parametrised through group mass ($M_{\rm group}$), and SFH parameters (Age, $\mu_{\rm SFH}$ and $\delta_{\rm SFH}$). The impact of environment on the spin parameter $\lambda_{R_e}$ has come under much scrutiny in recent years \citep[e.g.][]{Brough17, Greene17, Wang20, vandeSande21}, with evidence suggesting that environment plays a secondary role in setting the dynamics of galaxies, while stellar mass plays a more significant role. Using IFS data of galaxies at intermediate redshift in the COSMOS fields, \citet{MunozLopez24} found no dependence of stellar spin on environment. \citet{Croom24} recently found that it is age, rather than environment or stellar mass, that is most strongly correlated for stellar spin.

What is then the direct impact of environment, if any, in setting the dynamical properties of galaxies? Since spin is known to correlate with age \citep{Croom24}, star formation \citep[e.g.][]{Wang20} and stellar mass \citep[e.g.][]{vandeSande21}, we turn to more subtle dynamical measurements in the aim to detect any effect of environment on galaxy dynamics. \citet{DEugenio23b} have shown there is clear evolution towards higher integrated $H_4$ with decreasing redshift in massive galaxies, albeit with large scatter. This suggests an evolution of the central LOSVD of galaxies with cosmic time. In \citet{DEugenio23b}, $H_4$ values at low redshifts were interpreted as ``the outcome of accretion of gas-poor satellites''. Evolution of the shape of the LOSVD could arise through secular evolution \citep[e.g. the impact of bars][]{Bureau05,Iannuzzi15} or internal dynamical heating.
\citet{Bagge23} found differences in the kinemetric asymmetries $V_{\rm asym}$ of centrals and satellites. These studies suggest that higher-order kinematic moments and kinematic asymmetries may encode the subtle impact of environment on galaxy dynamics.

Here, we choose to address these questions using a mixture of simple correlation (\S\ref{sec:simplecorr}), principal component (\S\ref{sec:PCA}) and partial correlation (\S\ref{sec:pcorrs}) analyses to carefully isolate the impact of environment on our chosen dynamical parameters after accounting for other known factors (i.e. stellar age, SFH and stellar mass).
We confirm that known trends seen locally ($z\sim0$) between the spin parameter $\lambda_{R_e}$ and stellar mass and age are established in our $z\sim0.3$ MAGPI sample of galaxies (see Figures \ref{fig:corrs}, \ref{fig:allpcorrs} and Table \ref{table:corrs}). Indeed, as was identified in \citet{Croom24}, age most strongly correlates with $\lambda_{R_e}$ and is the only significant partial correlation identified for this dynamical parameter in our sample also.

Similarly, we see the trends already reported in \citet{Bagge23} and \citet{Bagge24} for SAMI and MAGPI between $V_{\rm asym}$ and stellar mass and age are also present in our selected sub-sample (Figure \ref{fig:corrs} and Table \ref{table:corrs}), though the trend with stellar age is no longer statistically significant once accounting for stellar mass (Figure \ref{fig:allpcorrs} and Table \ref{table:corrs}). In addition to these known trends, we find statistically significant trends between $\mu_{h_4}$ and $M_{\rm group}$ and stellar age (Figure \ref{fig:corrs} and Table \ref{table:corrs}). 

Of all considered dynamical parameters, $\mu_{h_4}$ is the only one that exhibits significant (albeit weak) partial correlations with environment ($M_{\rm group}$, see Figure \ref{fig:allpcorrs}). This indicates that $\mu_{h_4}$ encodes different properties than other considered dynamical parameters like $\lambda_{R_e}$ and $V_{\rm asym}$, which do not appreciably correlate with environment \citep[also see e.g. Figure \ref{fig:allpcorrs},][]{Greene17,Bagge23,Croom24}.

While there is no significant \textit{monotonic} (an assumption of the Spearman rank correlations methodology used in this work) trend detected with $\delta_{\rm SFH}$, we note that there is a dearth of galaxies in the upper right corners of the right-most panels of the top 2 rows in Figure \ref{fig:corrs} (i.e. $\delta_{\rm SFH}$ vs. $\rho_{V-h_3}$ and $\mu_{h_4}$). This suggests that galaxies in our sample that have had the most extended SFHs tend to have lower $\mu_{h_4}$ and have been able to retain a stronger $V-h_3$ anti-correlation (i.e. have more disc-like rotation). 
This indicates that these systems have continued to build their disc over an extended period through replenishment of fuel for star formation \citep[replenishment is required to sustain star formation, e.g.][]{Tacconi18} under conditions that have allowed them to retain and/or reform a disc.


Delving deeper into the principal component analysis described in \S\ref{sec:PCA} and shown in Figure \ref{fig:pca}, we find subtle hints of the covariance of the parametrised peak ($\mu_{\rm SFH}$) and extent ($\delta_{\rm SFH}$) of the SFH on $\mu_{h_4}$.
Figure \ref{fig:pca} illustrates how $\mu_{h_4}$ is represented primarily in principal components PC1, PC4 and PC7. The only parameter that is not represented in PC1 is $\delta_{\rm SFH}$, but it is represented at a low level as part of PC7, where $\lambda_{R_e}$ is not. This suggests that it may indeed be mainly overall age that co-varies with $\lambda_{R_e}$ rather than the details of the SFH. 

Our results show that while the environment may not have a significant measurable impact on the overall balance of rotation and random motions in galaxies as measured by $\lambda_{R_e}$, it does have a measurable impact on the more subtle shape of the local LOSVD as parametrised through $h_4$. This also is consistent with the concept that the higher-order kinematic moment $\mu_{h_4}$ may encode the presence of hot orbits in the wings of the LOSVD added through the cumulative impact of merging on galaxies as suggested by \citet[however noting the distinct methodology in computing an overall $h_4$ parameter used here as stated in \S\ref{sec:paramselect}]{DEugenio23a,DEugenio23b}. Similarly, \citet{Santucci23} find suggestive evidence that low mass galaxies in denser environments have a higher proportion of hot (over warm) stellar orbits even when accounting for mass, albeit without accounting for age. These lines of evidence suggest that the impact of cumulative mergers as measured through the proportion of stars on hotter orbits may indeed be more pronounced in dense environments, consistent with expectations from hierarchical assembly.

We have verified that $H_4$ and $\mu_{h_4}$ correlate in our sample (Figure \ref{fig:H4_muh4}), with some scatter at the lower values, usually associated with lower mass and fainter systems (consistent with Remus et al. in prep). Despite this correlation, the two parameters may yet encode different phenomena. In \ref{sec:appendix1}, we repeat our partial correlation analysis using $H_4$ and find that our conclusions are qualitatively unchanged, but note that there are proportionally more centrals included when using $H_4$, which does alter some of the sample-wide trends.

When considering satellites and centrals separately, we find that the environmental impact on $\mu_{h_4}$ (and indeed $H_4$) is evident only in satellite galaxies (Figures \ref{fig:allpcorrs} and \ref{fig:satcen}). Importantly, we find that the dynamics of satellites and centrals as measured using $\lambda_{R_e}$ and $\mu_{h_4}$ at $z\sim0.3$ are distinct, with the former behaving in agreement with local trends discussed in \citet{Croom24}. 

The fact that the partial correlation of $\mu_{h_4}$ with group mass disappears when considering only central galaxies in Figure \ref{fig:satcen}, suggests that the shape of the LOSVD in galaxies that dominate their environment mainly reflects the time at which the bulk of their stars were formed. 
This may suggest that the mass of the group they dominate is less relevant to the LOSVD shape than how long they have dominated their environment.
The trend between $h_4$ and environment in satellites being more significant in contrast to centrals suggests that galaxies that do not dominate their environments have distinct orbit families. At fixed stellar age, satellites in less massive groups exhibit a lower proportion of stars on hotter orbits in the LOSVD than those in more massive groups. 

\section{Conclusions}\label{sec:conclusions}
We explore the simultaneous impact of stellar mass, environment and successive generations of stars on the dynamical properties of galaxies in a sample of 77 galaxies (including 41 centrals, 34 satellites and 2 isolated targets) at intermediate redshift ($z\sim0.3$) in the MAGPI Survey. We make use of a range of tools, including partial correlations and principal component analyses to delineate contributing factors and isolate the impact of environment. In particular, we explicitly account for factors known to correlate with dynamics such as age, the width and peak of the star formation histories and stellar masses. We select 4 dynamical parameters:
\begin{enumerate}
\item the Spearman correlation coefficient of the anti-correlation between $V$ and $h_3$ within one effective radius ($\rho_{V-h_3}$);
\item the mean $h_4$ within one effective radius ($\mu_{h_4}$);
\item the kinematic asymmetry measured on the stellar kinematic maps using kinemetry at $1R_e$ ($V_{\rm asym}$); and
\item the traditional spin parameter $\lambda_{R_e}$ measured within one effective radius.
\end{enumerate}

Our main conclusions are:
\begin{itemize}
\item The dynamical parameter $\mu_{h_4}$ is the only considered dynamical parameter found to have a significant residual correlation with environment as parametrised by $M_{\rm group}$ (Figure \ref{fig:allpcorrs}) after accounting for stellar mass and age. This suggests that the shape of the LOSVD in galaxies (particularly satellites) measurably varies with group mass.
\item Satellite and central galaxies exhibit distinct trends, suggesting they are dynamically distinct classes (Figures \ref{fig:allpcorrs} and \ref{fig:satcen}).
\item We confirm that variations in the spin parameter $\lambda_{R_e}$ are primarily correlated with age (also at $z\sim0.3$, see Figure \ref{fig:allpcorrs}).
\end{itemize}

Future work will focus on comparisons with hydrodynamical simulations using the MAGPI theoretical dataset and local IFS surveys to confirm the trends seen here and identify the main physical processes that are involved in setting the dynamical state of galaxies. 
We note that the current MAGPI sample does not include the highest density environments such as massive clusters. Our results will be worth revisiting once consistently analysed cluster data are available.

\begin{acknowledgement}
AP acknowledges support from the Hintze Family Charitable Foundation.

Based on observations collected at the European Organisation for Astronomical Research in the Southern Hemisphere under ESO program 1104.B-0536. We wish to thank the ESO staff, and in particular the staff at Paranal Observatory, for carrying out the MAGPI observations. 

MAGPI targets were selected from GAMA. GAMA is a joint European-Australasian project based around a spectroscopic campaign using the Anglo-Australian Telescope. GAMA is funded by the STFC (UK), the ARC (Australia), the AAO, and the participating institutions. GAMA photometry is based on observations made with ESO Telescopes at the La Silla Paranal Observatory under programme ID 179.A-2004, ID 177.A-3016.

This work makes use of colour scales chosen from \citet{CMASHER2020}. 
\end{acknowledgement}

\paragraph{Funding Statement}

CF is the recipient of an Australian Research Council Future Fellowship (project number FT210100168) funded by the Australian Government.
CL, JTM and CF are the recipients of ARC Discovery Project DP210101945.
Part of this research was conducted by the Australian Research Council Centre of Excellence for All Sky Astrophysics in 3 Dimensions (ASTRO 3D), through project number CE170100013.
FDE acknowledges support by the Science and Technology Facilities Council (STFC), by the ERC through Advanced Grant 695671 ``QUENCH'', and by the UKRI Frontier Research grant RISEandFALL.
LMV acknowledges support by the German Academic Scholarship Foundation (Studienstiftung des deutschen Volkes) and the Marianne-Plehn-Program of the Elite Network of Bavaria.

\paragraph{Competing Interests}

None.

\paragraph{Data Availability Statement}

The MAGPI raw data (and a basic data reduction) are available through the \href{http://archive.eso.org/cms.html}{ESO Science Archive Facility}.

\bibliography{biblio}

\appendix

\section{Calibration of the ppxf bias value}\label{sec:ppxfbias}

\begin{figure}
\includegraphics[width=8.5cm]{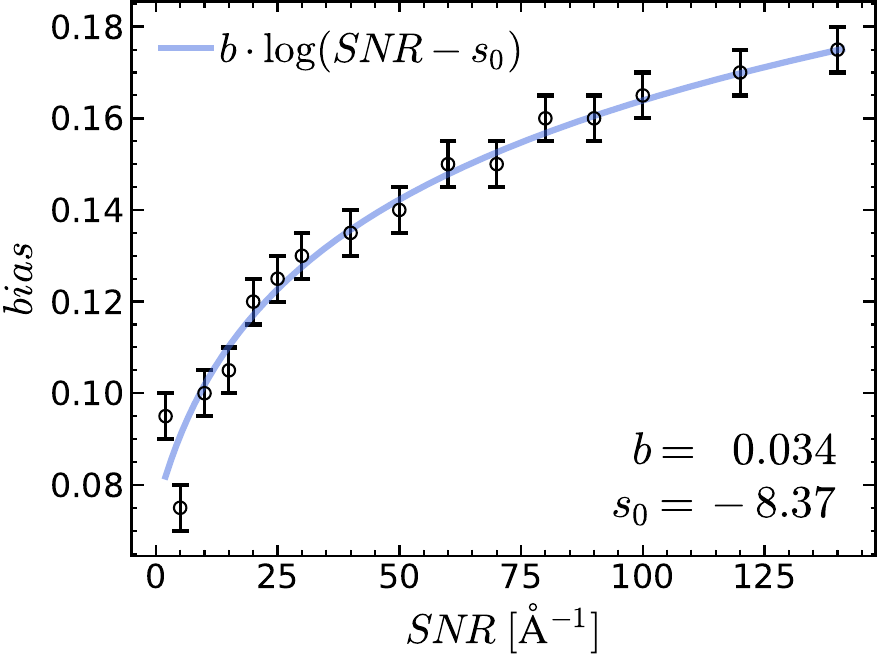}
\caption{Calibration of the {\sc pPXF} bias keyword vs  $S/N$ for the MAGPI spectra. The circles (errorbars) ar e the fiducial value and the 16--84\textsuperscript{th} percentile range. The solid blue line is the best-ﬁt to the data.}
\label{fig:bias.keyword}
\end{figure}

For spectra with $S/N\lesssim15~$\AA, the measured LOSVD can be significantly biased to non-Gaussian solutions \citep{Cappellari04}. To minimise over-fitting, we use the `bias' feature of {\sc pPXF}, which `penalises' the minimum-$\chi^2$ solution for non-Gaussian deviations using a a penalisation factor set by the keyword \texttt{bias}. The precise value of \texttt{bias} requires calibration tailored to the characteristics of the data, i.e. spectral resolution, wavelength range, and $S/N$ \citep[e.g.,][]{vandeSande17a}. Following the approach of the SAMI Galaxy Survey \citep[see][]{vandeSande17a}, we use random-noise realisations of the best-fit spectrum of two high-$S/N$ MAGPI observations, galaxies MAGPI1202197197 and MAGPI2301177186, chosen to represent a star-forming and a quiescent galaxy, respectively.
The spectra were created from the linear combination of the best-fit IndoUS templates, oversampling the spectra by a factor of three \citep{Cappellari04}. We then convolve the spectra with the instrument resolution, and with a range of trial LOSVDs. The latter span 100 velocities $-150<v<150$~\kms, velocity dispersions $5<\sigma<350$~\kms in steps of 5~\kms, and $h_3 = 0.1 = h_4$. Finally, we add random Gaussian noise, with a range of $S/N$  values spanning 2--140~\AA$^{-1}$.
The resulting range of spectra were fit with {\sc pPXF}, using a range of 80 values of \texttt{bias} keyword spanning uniformly the range 0--0.4. For each LOSVD, and at each S/N value, the optimal \texttt{bias} keyword minimises the bias of the solution, and the uncertainty on the recovered parameters following the methodology introduced in \citet{vandeSande17a}.
Figure~\ref{fig:bias.keyword} shows how the distribution of optimal \texttt{bias} keyword varies as a function of the input $S/N$.
We model these values alternatively as a 2\textsuperscript{nd}-order polynomial in $S/N$ and as a logarithm of $S/N$, finding that the favoured (lowest reduced-$\chi^2$) model is
\begin{equation}\label{eq:bias}
   \texttt{bias} = 0.034 \ln(S/N + 8.37)
\end{equation}
This function is then used when setting \texttt{bias} for each galaxy and at each spaxel.

\section{\texorpdfstring{$H_4$}{H4} vs. \texorpdfstring{$\mu_{h_4}$}{muh4} comparison}\label{sec:appendix1}


Our choice of taking the weighted mean of $h_4$ available spaxels with signal-to-noise $>15$ within $1R_e$ instead of measuring $H_4$ on the integrated spectrum within $1R_e$ is arguably controversial. This was done to mitigate the potential impact of high rotation in artificially altering the LOSVD and hence skewing our results for $H_4$. In this section, we rerun the relevant partial correlation analysis to ensure that our choice of $\mu_{h_4}$ over $H_4$ does not lead to spurious conclusions. The main goal here is to ascertain that our main conclusions are robust against the choice of parameter. We have also tested that our results are similarly robust against different choices of signal-to-noise thresholds (not shown).

\begin{figure}
\includegraphics[width=8.5cm]{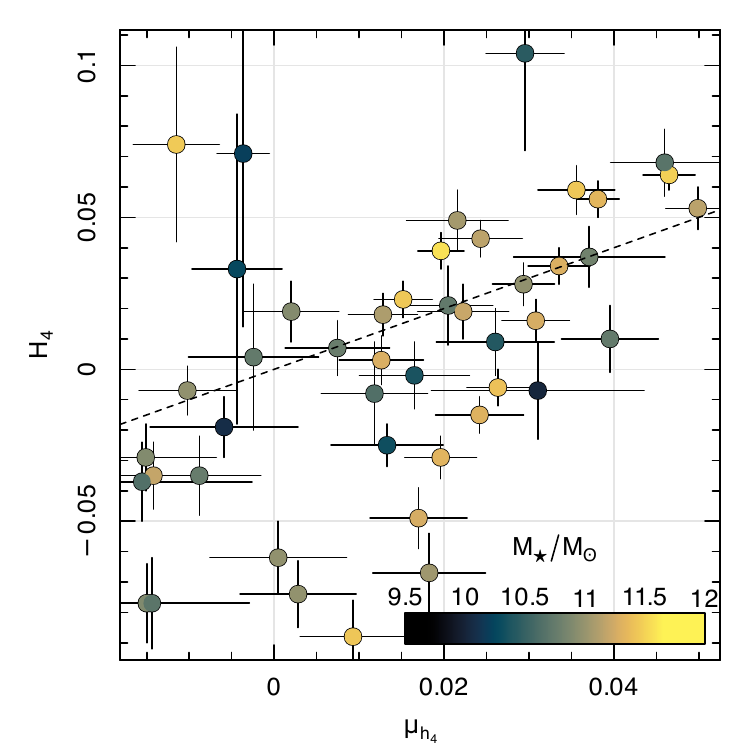}
\caption{Comparison of weighted average $h_4$ ($\mu_{h_4}$) used in this work and the $H_4$ measured on the integrated $1R_e$ aperture spectrum as per \citet{DEugenio23b} colour-coded by stellar mass. There is a statistically significant (Spearman rank coefficient of $\rho=0.52$ with $p$-value of 0.0002) correlation between the two parameters that scatters about the one-to-one (dashed line), with increasing scatter towards low values of $\mu_{h_4}$.}
\label{fig:H4_muh4}
\end{figure}

We begin by confirming that $H_4$ and $\mu_{h_4}$ correlate in Figure \ref{fig:H4_muh4}, suggesting they may trace similar physical processes. Although we note that at lower masses and low $\mu_{h_4}$, there is increased scatter in $H_4$. This may either be a feature of real differences in the physical processes probed by each parameter or attributed to the inclusion of lower signal-to-noise spaxels in the integrated spectrum for $H_4$ or to low number statistics plaguing $\mu_{h_4}$ in the limit where few spaxels meet the signal-to-noise threshold. This higher scatter needs to be borne in mind as it may weaken other trends with either parameter and our conclusions.

\begin{figure}
\includegraphics[width=8.5cm]{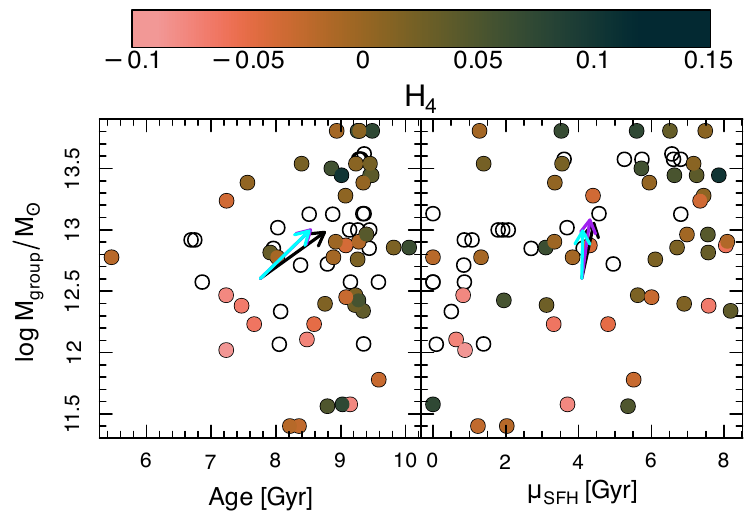}
\caption{Same as the top row of Figure \ref{fig:allpcorrs}, but for $H_4$ measured as per \citet{DEugenio23b}. Partial correlation Spearman rank coefficients and $p$-values are stated in Table \ref{table:pcorrs_H4}.}
\label{fig:frh4pcorrs}
\end{figure}

\begin{figure}
\includegraphics[width=8.5cm]{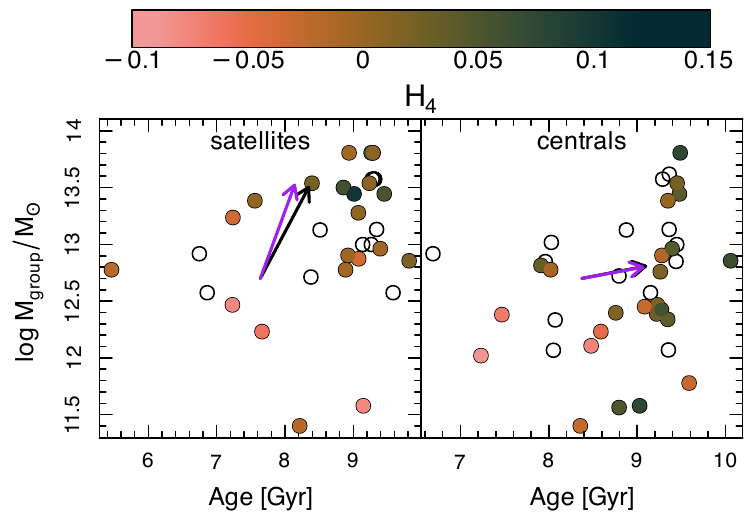}
\caption{Same as the top row of Figure \ref{fig:satcen} but for $H_4$ measured as per \citet{DEugenio23b}. Using $H_4$, the differences between satellites and centrals are even more marked than for $\mu_{h_4}$ (see Figure \ref{fig:satcen}). Partial correlation Spearman rank coefficients and $p$-values are stated in Table \ref{table:pcorrs_H4}.}
\label{fig:frsatcen}
\end{figure}

\begin{table*}
\caption{Compiled Spearman rank partial correlation coefficients $\rho$ and respective confidence ($p$-values) for $H_4$. Significant correlations (i.e. $p\le 0.02$) are highlighted in bold. Thus, $x = {\rm Age}$ or $\mu_{\rm SFH}$, $y=M_{\rm group}$, and $\textbf{Z}$ a subset of $\{H_4, M_\star, \mu_{\rm SFH}\}$ (as labelled in column 1) as per Eq. \ref{eq:pcorr}. The dynamical parameter $H_4$ is compared with group mass ($M_{\rm group}$, column 2), mass weighted stellar age (Age, column 3), the lookback time of the peak of the star formation history ($\mu_{\rm SFH}$, column 4); and stellar mass ($M_\star$, column 5). Results are broadly consistent with that found for $\mu_{h_4}$ as listed in Table \ref{table:pcorrs}, but see text for a detailed discussion of the contrasts.}
\input{table_pcorrs_H4}
\label{table:pcorrs_H4}
\end{table*}

We next repeat the partial correlation analysis using $H_4$ instead of $\mu_{h_4}$. Results are shown in Figure \ref{fig:frh4pcorrs} for the whole sample and in Figure \ref{fig:frsatcen} for satellites and centrals separately. Correlation coefficients and respective p-values are shown in Table \ref{table:pcorrs_H4}. There are a few contrasts worth highlighting that emerge when comparing Tables \ref{table:pcorrs} and \ref{table:pcorrs_H4}.

The partial correlations of $H_4$ with $M_{\rm group}$ are generally less significant (lower $p$-values than those with $\mu_{h_4}$ possibly due to the reduced sample size). The partial correlation of centrals with stellar age becomes significant when using $H_4$, it was not significant when using $\mu_{h_4}$, possibly due to the different ratio of centrals to satellites in the samples used. The other difference is that the partial correlation with group mass when accounting for all other parameters is marginal (rather than significant) when using $H_4$.

All other significant partial correlations are present in both $H_4$ and $\mu_{h_4}$. We thus conclude that our choice of $\mu_{h_4}$ does not significantly alter the conclusions described in this work.

\end{document}

%% file: table_spearman.tex
\centering
\resizebox{\columnwidth}{!}{
\begin{tabular}{| l || c c c | c c c | c c c | c c c | c c c |}
\hline
\backslashbox{$y$}{$x$} & \multicolumn{3}{|c|}{$M_\star$} & 
  \multicolumn{3}{|c|}{$M_{\rm group}$} & \multicolumn{3}{|c|}{Age} 
  & \multicolumn{3}{|c|}{$\mu_{SFH}$}& \multicolumn{3}{|c|}{$\delta_{SFH}$}\\
& \multicolumn{3}{|c|}{(1)} & \multicolumn{3}{|c|}{(2)} & \multicolumn{3}{|c|}{(3)} & 
  \multicolumn{3}{|c|}{(4)} & \multicolumn{3}{|c|}{(5)}\\
& $\rho$ & $p$ & $N_{\rm CC}$ & $\rho$ & $p$ & $N_{\rm CC}$ & $\rho$ & 
      $p$ & $N_{\rm CC}$ & $\rho$ & $p$ & $N_{\rm CC}$ & $\rho$ & $p$ & $N_{\rm CC}$\\
\hline
$\rho_{V-h_3}$ & 0.18$\pm$0.16 & 0.13 & 72 & -0.021$\pm$0.018 & 0.87 & 70 & 0.094$\pm$0.53 & 0.43 & 72 & 0.017$\pm$0.014 & 0.88 & 72 & 0.055$\pm$0.14 & 0.65 & 72 \\
$\mu_{h_4}$ & 0.26$\pm$0.14 & 0.035 & 68 & \textbf{0.38$\pm$0.12} & \textbf{0.0017} & 67 & \textbf{0.48$\pm$0.11} & \textbf{0.00003} & 68 & 0.24$\pm$0.14 & 0.049 & 68 & -0.18$\pm$0.17 & 0.14 & 68 \\
$V_{\rm asym}$ & \textbf{-0.33$\pm$0.14} & \textbf{0.013} & 55 & -0.021$\pm$0.018 & 0.88 & 55 & \textbf{-0.38$\pm$0.14} & \textbf{0.0047} & 55 & -0.15$\pm$0.26 & 0.28 & 55 & 0.16$\pm$0.21 & 0.23 & 55 \\
$\lambda_{R_e}$ & -0.26$\pm$0.13 & 0.029 & 73 & -0.22$\pm$0.15 & 0.070 & 71 & \textbf{-0.55$\pm$0.10} & \textbf{0.0000004} & 73 & -0.18$\pm$0.15 & 0.12 & 73 & 0.040$\pm$0.062 & 0.74 & 73 \\
\hline
\end{tabular}}

%% file: table_pcorrs.tex
\begin{tabular}{| l || c c | c c | c c | c c | c |}
\hline
Fitted parameters & \multicolumn{2}{|c|}{$M_{\rm group}$} & \multicolumn{2}{|c|}{Age} & 
      \multicolumn{2}{|c|}{$\mu_{\rm SFH}$} & \multicolumn{2}{|c|}{$M_\star$} & $N_{\rm CC}$\\
($x,y,\textbf{Z}$) & \multicolumn{2}{|c|}{} & \multicolumn{2}{|c|}{} & 
     \multicolumn{2}{|c|}{} & \multicolumn{2}{|c|}{} &\\
(1) & \multicolumn{2}{|c|}{(2)} & \multicolumn{2}{|c|}{(3)} & \multicolumn{2}{|c|}{(4)} 
      & \multicolumn{2}{|c|}{(5)} & (6) \\
& $\rho$ & $p$ & $\rho$ & $p$ & $\rho$ & $p$ & $\rho$ & $p$ & \\
\hline\hline
$x={\rm Age},y=M_{\rm group},\textbf{Z}=\{\mu_{h_4}\}$ & \textbf{0.29} & \textbf{0.019} & \textbf{0.39} & \textbf{0.0015} & -- & -- & -- & -- & 66 \\
$x=\mu_{\rm SFH},y=M_{\rm group},\textbf{Z}=\{\mu_{h_4}\}$ & \textbf{0.32} & \textbf{0.0092} & -- & -- & 0.15 & 0.24 & -- & -- & 66 \\
$x={\rm Age},y=M_{\rm group},\textbf{Z}=\{\mu_{h_4},M_\star\}$ & \textbf{0.30} & \textbf{0.016} & \textbf{0.31} & \textbf{0.013} & -- & -- & 0.14 & 0.27 & 65 \\
$x=\mu_{\rm SFH},y=M_{\rm group},\textbf{Z}=\{\mu_{h_4},M_\star\}$ & \textbf{0.34} & \textbf{0.0060} & -- & -- & 0.086 & 0.50 & 0.24 & 0.060 & 66 \\
$x={\rm Age},y=M_{\rm group},\textbf{Z}=\{\mu_{h_4},\mu_{\rm SFH},M_\star\}$ & 0.28 & 0.027 & \textbf{0.31} & \textbf{0.016} & 0.043 & 0.74 & 0.13 & 0.31 & 65 \\
$x={\rm Age},y=M_{\rm group},\textbf{Z}=\{\mu_{h_4},\mu_{\rm SFH},M_\star\}$ (centrals) & -0.061 & 0.74 & 0.32 & 0.073 & 0.27 & 0.14 & 0.23 & 0.20 & 35 \\
$x={\rm Age},y=M_{\rm group},\textbf{Z}=\{\mu_{h_4},\mu_{\rm SFH},M_\star\}$ (satellites) & \textbf{0.62} & \textbf{0.00052} & 0.29 & 0.14 & -0.17 & 0.39 & 0.057 & 0.78 & 30 \\
\hline\hline
$x={\rm Age},y=M_{\rm group},\textbf{Z}=\{V_{\rm asym}\}$ & 0.099 & 0.48 & \textbf{-0.39} & \textbf{0.0042} & -- & -- & -- & -- & 54 \\
$x=\mu_{\rm SFH},y=M_{\rm group},\textbf{Z}=\{V_{\rm asym}\}$ & 0.016 & 0.91 & -- & -- & -0.16 & 0.25 & -- & -- & 54 \\
$x={\rm Age},y=M_{\rm group},\textbf{Z}=\{V_{\rm asym},M_\star\}$ & 0.091 & 0.53 & -0.29 & 0.040 & -- & -- & -0.24 & 0.09 & 53 \\
$x=\mu_{\rm SFH},y=M_{\rm group},\textbf{Z}=\{V_{\rm asym},M_\star\}$ & 0.021 & 0.89 & -- & -- & -0.064 & 0.65 & -0.32 & 0.021 & 54 \\
$x={\rm Age},y=M_{\rm group},\textbf{Z}=\{V_{\rm asym},\mu_{\rm SFH},M_\star\}$ & 0.096 & 0.51 & -0.29 & 0.044 & -0.038 & 0.79 & -0.22 & 0.12 & 53 \\
$x={\rm Age},y=M_{\rm group},\textbf{Z}=\{V_{\rm asym},\mu_{\rm SFH},M_\star\}$ (centrals) & 0.093 & 0.64 & -0.15 & 0.45 & 0.071 & 0.72 & -0.25 & 0.21 & 30 \\
$x={\rm Age},y=M_{\rm group},\textbf{Z}=\{V_{\rm asym},\mu_{\rm SFH},M_\star\}$ (satellites) & 0.11 & 0.64 & -0.45 & 0.046 & -0.23 & 0.33 & -0.21 & 0.38 & 23 \\
\hline\hline
$x={\rm Age},y=M_{\rm group},\textbf{Z}=\{\lambda_{R_e}\}$ & -0.0096 & 0.94 & \textbf{-0.55} & \textbf{0.000001} & -- & -- & -- & -- & 70 \\
$x=\mu_{\rm SFH},y=M_{\rm group},\textbf{Z}=\{\lambda_{R_e}\}$ & -0.15 & 0.21 & -- & -- & -0.16 & 0.18 & -- & -- & 70 \\
$x={\rm Age},y=M_{\rm group},\textbf{Z}=\{\lambda_{R_e},M_\star\}$ & -0.0041 & 0.97 & \textbf{-0.50} & \textbf{0.00002} & -- & -- & -0.030 & 0.81 & 69 \\
$x=\mu_{\rm SFH},y=M_{\rm group},\textbf{Z}=\{\lambda_{R_e},M_\star\}$ & -0.17 & 0.16 & -- & -- & -0.096 & 0.44 & -0.25 & 0.038 & 70 \\
$x={\rm Age},y=M_{\rm group},\textbf{Z}=\{\lambda_{R_e},\mu_{\rm SFH},M_\star\}$ & 0.0061 & 0.96 & \textbf{-0.49} & \textbf{0.00003} & -0.039 & 0.76 & -0.022 & 0.86 & 69 \\
$x={\rm Age},y=M_{\rm group},\textbf{Z}=\{\lambda_{R_e},\mu_{\rm SFH},M_\star\}$ (centrals) & 0.036 & 0.84 & \textbf{-0.54} & \textbf{0.0013} & -0.092 & 0.61 & -0.059 & 0.74 & 36 \\
$x={\rm Age},y=M_{\rm group},\textbf{Z}=\{\lambda_{R_e},\mu_{\rm SFH},M_\star\}$ (satellites) & -0.10 & 0.59 & -0.33 & 0.08 & -0.069 & 0.72 & 0.014 & 0.94 & 33 \\
\hline
\end{tabular}

%% file: table_pcorrs_H4.tex
\begin{tabular}{| l || c c | c c | c c | c c | c |}
\hline
Fitted parameters & \multicolumn{2}{|c|}{$M_{\rm group}$} & \multicolumn{2}{|c|}{Age} & 
      \multicolumn{2}{|c|}{$\mu_{\rm SFH}$} & \multicolumn{2}{|c|}{$M_\star$} & $N_{\rm CC}$\\
($x,y,\textbf{Z}$) & \multicolumn{2}{|c|}{} & \multicolumn{2}{|c|}{} & 
     \multicolumn{2}{|c|}{} & \multicolumn{2}{|c|}{} &\\
(1) & \multicolumn{2}{|c|}{(2)} & \multicolumn{2}{|c|}{(3)} & \multicolumn{2}{|c|}{(4)} 
      & \multicolumn{2}{|c|}{(5)} & (6) \\
& $\rho$ & $p$ & $\rho$ & $p$ & $\rho$ & $p$ & $\rho$ & $p$ & \\
\hline\hline
$x={\rm Age},y=M_{\rm group},\textbf{Z}=\{H_4\}$ & 0.31 & 0.037 & \textbf{0.43} & \textbf{0.0031} & -- & -- & -- & -- & 46 \\
$x=\mu_{\rm SFH},y=M_{\rm group},\textbf{Z}=\{H_4\}$ & \textbf{0.37} & \textbf{0.0099} & -- & -- & 0.077 & 0.61 & -- & -- & 48 \\
$x={\rm Age},y=M_{\rm group},\textbf{Z}=\{H_4,M_\star\}$ & 0.33 & 0.029 & 0.34 & 0.025 & -- & -- & 0.11 & 0.46 & 46 \\
$x=\mu_{\rm SFH},y=M_{\rm group},\textbf{Z}=\{H_4,M_\star\}$ & \textbf{0.4} & \textbf{0.0064} & -- & -- & 0.052 & 0.73 & 0.27 & 0.065 & 48 \\
$x={\rm Age},y=M_{\rm group},\textbf{Z}=\{H_4,\mu_{\rm SFH},M_\star\}$ & 0.32 & 0.036 & 0.33 & 0.032 & 0.0057 & 0.97 & 0.11 & 0.46 & 46 \\
$x={\rm Age},y=M_{\rm group},\textbf{Z}=\{H_4,\mu_{\rm SFH},M_\star\}$ (centrals) & 0.11 & 0.64 & 0.42 & 0.060 & -0.085 & 0.72 & 0.013 & 0.95 & 24 \\
$x={\rm Age},y=M_{\rm group},\textbf{Z}=\{H_4,\mu_{\rm SFH},M_\star\}$ (satellites) & \textbf{0.66} & \textbf{0.0022} & 0.16 & 0.50 & 0.14 & 0.57 & 0.15 & 0.53 & 22 \\
\hline
\end{tabular}